\documentclass[12pt,a4paper]{article}
\usepackage{comment}
\usepackage{amssymb}
\usepackage{amsmath}
\usepackage{amsfonts}
\usepackage{enumerate}
\usepackage{float}
\usepackage{natbib}
\usepackage{verbatim}
\usepackage{graphicx}
\usepackage{lscape}
\usepackage[autostyle]{csquotes}
\usepackage{bbm}
\usepackage[dvipsnames]{xcolor}
\usepackage[a4paper,bindingoffset=0.2in,left=1in,right=1in,top=1in,bottom=1in,footskip=.5in]{geometry}
\usepackage{mathrsfs} 
\usepackage{xfrac}

\usepackage{multirow}
\usepackage{longtable}
\usepackage{float}
\usepackage{esvect}
\usepackage[T1]{fontenc}
\usepackage{amsmath}
\usepackage{setspace}
\usepackage[utf8]{inputenc}
\usepackage{xcolor}
\usepackage{mathrsfs} 
\usepackage[autostyle]{csquotes}
\usepackage[flushleft]{threeparttable}

\graphicspath{ {images/} }   

\DeclareMathAlphabet{\pazocal}{OMS}{zplm}{m}{n}

\begin{document}

\title{Forecasting With Factor-Augmented Quantile Autoregressions: A Model Averaging Approach\thanks{%
I am greatly indebted to Valentina Corradi and Vasco Gabriel for their constant guidance throughout this project. I would also like to thank Joao Santos Silva, Ricardo P. C. Nunes, Ekaterina Oparina, Ayden Higgins and the participants of the Econometrics Workshop at the University of Surrey for helpful comments and discussions.The author also thanks the South East Network for Social Sciences for financial support.}}
\author{Anthoulla Phella{\small \thanks{\textit{E-mail address}:
a.phella@surrey.ac.uk}} \\
%EndAName
University of Surrey}
\date{\today}
\maketitle

\begin{abstract}
This paper considers forecasts of the growth and inflation distributions of the United Kingdom with factor-augmented quantile autoregressions under a model averaging framework. We investigate model combinations across models using weights that minimise the Akaike Information Criterion (AIC), the Bayesian Information Criterion (BIC), the Quantile Regression Information Criterion (QRIC) as well as the leave-one-out cross validation criterion. The unobserved factors are estimated by principal components of a large panel with N predictors over T periods under a recursive estimation scheme.  We apply the aforementioned methods to the UK GDP growth and CPI inflation rate. We find that, on average, for GDP growth, in terms of coverage and final prediction error, the equal weights or the weights obtained by the AIC and BIC perform equally well but are outperformed by the QRIC and the Jackknife approach on the majority of the quantiles of interest. In contrast, the naive QAR(1) model of inflation outperforms all model averaging methodologies.\\


\end{abstract}

\vspace{2cm}

\setcounter{page}{1}\newpage
\section{Introduction}
Model averaging is an alternative to model selection which has seen an increase in popularity over the recent years. In a model selection approach, the researcher attempts to find a single best model for a given purpose, while ignoring all the information in other models. Under the model averaging approach, however, the researcher combines information from all competing models by obtaining a weighted average of the competing models' estimators. This can be seen as an insurance against selecting a very poor model, as it allows the researcher to diversify and account for model uncertainty, thus enabling an improvement in out-of-sample performance. This is because, as argued by inter alia \citet{hendry_pooling_2004, wallis_combining_2005}, even a simple combination of forecasts with equal weights can never produce a forecast that is worse than the worst individual forecast. Frequentist model averaging (FMA), despite having a shorter history in economics than Bayesian Model Averaging (BMA), has started to receive a lot of attention in the past decade or so, as, in contrast with the BMA, it requires no priors and the corresponding estimators are totally determined by the data \citep{moral-benito_model_2015} .\\

At the same time, the recent availability of large datasets has generated interest in models with many possible predictors. Factor-augmented regressions, in particular, have been proven to forecast a particular series relatively well compared to the original predictors, with the benefit of significant dimension reduction \citep{stock_forecasting_2002}. However, factors are usually determined and ordered by their importance in driving the co-variability of many predictors, which may not necessarily be consistent with their forecast power over a particular series of interest. Model specification is therefore necessary in order to determine which factors should be included in the forecast regression, in addition to specifying the number of lags of all independent variables. \\

In this paper, we consider different model averaging methodologies for the combination of different quantile autoregressive (QAR) models, in an attempt to best forecast the quantiles of inflation and growth. \citet{kapetanios_forecast_2008} provide compelling reasons for using model averaging for the purpose of forecasting, in order to address the problem of model uncertainty. Combinations of forecasts often outperform individual forecasts, as models may be incomplete in different ways, thus averaging can offset different biases present in each model \citep{granger_can_1996, granger_spurious_1974}. Nevertheless, this prior empirical work on forecast combinations of inflation focuses on point estimates for the conditional mean of inflation \citep{kapetanios_forecast_2008} and such point forecasts ignore the risks and uncertainty around the central forecast.  In the case of growth, for example, such forecasts may paint an overly optimistic picture of the state of the economy \citep{adrian_vulnerable_2019}.  Distribution or quantile forecasts, on the other hand, provide a more complete picture of the conditional dependence structure of the variables examined and allow us to forecast higher moments. Predicting several conditional quantiles that can characterise the entire distribution of future growth and inflation can be integral in order to properly assess growth vulnerability and inflation stability \citep{adrian_vulnerable_2019, manzan_are_2013}. \\

The intersection of latent factors with quantile regression models is fairly recent.  \citet{ando_quantile_2011} have considered a quantile regression model with factor-augmented predictors, whose effect is allowed to vary across the different quantiles.  More recently, \citet{ando_quantile_2020} introduced a new procedure for analysing the quantile co-movement of a large number of time series based on a large scale panel data model with factor structures. In their study, the latent factors are allowed to vary across the different quantiles of the variables from which they are extracted and, as such, their model is a quantile factor model. Similarly, \citet{chen_quantile_2019} estimate scale-shifting factors and quantile dependent loadings, thus factors may shift characteristics (moments or quantiles) of the distribution of the set of directly observable measures, other than its mean, and factor loadings are allowed to vary across the distributional characteristics of each variable.  \citet{phella_consistent_2020} contributed to this relevant literature by using mean shifting factors as a method of dimension reduction and use these latent factors as additional regressors in quantile autoregressive models. Results showed that the distribution of CPI inflation is best modelled parametrically with a quantile autoregression of order 1, while the distribution of GDP growth is best modelled as a factor-augmented quantile autoregression and such latent factors impact certain quantiles differently. Therefore, our pool of candidate models in the forecasting exercise in this paper will include multiple lag QARs, factor augmented QARs or QARs augmented with targeted macroeconomic variables.\\

Nevertheless, in order for the model averaging approach to outperform the model selection, weights assigned to each model need to be correctly chosen and how such weights are chosen can have a significant impact on forecast performance. \citet{buckland_model_1997} and \citet{burnham_model_2002} construct model averaging weights based on the values of the Akaike information criterion(AIC) and the Bayesian Information Criterion (BIC). In a seminal article, \citet{hansen_least_2007} proposed that weights in least squares model averaging should be chosen over a discrete set by minimizing a Mallow's criterion, as such an estimator is asymptotically optimal in terms of optimising the mean squared error. Meanwhile, \citet{hansen_jackknife_2012} propose a Jackknife model averaging (JMA) approach for least squares regression where weights are selected by minimising a leave-one-out cross-validation criterion function, which was extended for models with dependent data by \citet{zhang_model_2013}. More recently, \citet{cheng_forecasting_2015} demonstrated that both the Mallows and leave-h-out cross-validationn criteria remain valid in factor-augmented regression forecasts, as the factor estimation error is negligible, without any restrictions on the relation between N and T.\\

\citet{lu_jackknife_2015} extended the JMA of \citet{hansen_least_2007} to the quantile regression framework and also proposed a Mallows-type information criterion for QR model averaging, the Quantile Regression Information Criterion (QRIC), which has a computational advantage over the Jackknife approach. We therefore employ this multitude of weighting methodologies over the competing models. We utilise the AIC and the BIC, which use exponential weighting and are often referred to as \enquote{smoothed} averaging \citep{buckland_model_1997}, the Quantile Regression Information Criterion (QRIC) and the Jackknife weighting method, as those are outlined in \citet{lu_jackknife_2015}. We also choose to include in our forecast performance comparison the quantile autoregressive model of order 1, QAR(1), as a naive benchmark model similar to the one proposed in \citet{pasaogullari_simple_2010} and \citet{pasaogullari_simple_2010}, as well as a full model which includes all the possible predictors under consideration.\\

We employ these methodologies to determine which model averaging weighting choice is the best for forecast performance, in terms of coverage and final prediction error when predicting one-quarter-ahead GDP growth and CPI inflation for the United Kingdom.\footnote{Along with the coverage rates, we also consider an interval score in order to obtain information regarding the magnitude of violations when they take place.} We specifically employ these methodologies within our estimation sample only, in order to assign fixed weights to the competing models and then use these corresponding weights to obtain the averaged model and produce forecasts, which are in turn evaluated by the aforementioned performance measures. Our results demonstrate that, on average, the equal weights or the weights obtained by the AIC and BIC perform equally well, but are somehow outperformed by the QRIC and the Jackknife approach on the majority of the quantiles of interest of GDP growth, both in terms of coverage and final prediction error. Furthermore, the QRIC and Jackknife model averages ourperform the full model where all available information has been used.  On the other hand, the Quantile Autoregression of order 1 of CPI inflation, a model that is often chosen as the best predictor in forecasts of the average value, outperforms all model averaged methodologies. \\

The remainder of the paper is organised as follows. In Section 2, we outline the framework, present the range of models we consider and describe the model averaging methodologies.  Section 3 examines the forecast evaluations with respect to the two forecast performance measures. Concluding remarks are given in Section 4 and information regarding mnemonics and the competing models are referred to an appendix.\\

\vspace{0.5cm}

\section{The Framework}
\subsection{Quantile Autoregression Model Averaging}
We begin by outlining the factor model used in the sequel. Let
\begin{align}
X_t = \Lambda_t F_t + e_t  \quad, \label{factor extraction 2}
\end{align}\\
where  $X_t$ is an $N \times 1$ vector of observable variables characterising the economy, $\Lambda_t$ is an $N \times k$ matrix of factor loadings, $F_t$ is a $k \times 1$ vector of the $k$ latent common factors and $e_t$ is an $N \times 1$ vector of idiosyncratic disturbances. The errors are allowed to be both serially and (weakly) cross sectionally correlated.\\

As in \citet{stock_forecasting_2002}, factors are extracted via the principle components approach and the estimated factors and estimated factor loadings are defined as:
\begin{align*}
(\hat{F},\hat{\Lambda})=\arg \min_{F,\Lambda} \frac{1}{NT}\sum_{i=1}^N \sum_{t=1}^T (X_{it}-\Lambda_{i}F_{t}).
\end{align*}

The resulting principal components estimator of $F$ is then $\hat{F}=\frac{X'\hat{\Lambda}}{N}$, where $\hat{\Lambda}$ is set equal to the eigenvectors of $X'X$ corresponding to its $k$ largest eigenvalues. In the remainder of this paper, the number of factors $k$ would remain fixed and can be estimated using the information criteria outlined in \citet{bai_determining_2002}, who take into account the sample size both in the cross-section and time-series dimensions.\\ 

Ideally, we would like to include all the macroeconomic variables in $X_t$ as additional regressors in the quantile autoregressive model,  however, due to the curse of dimensionality, we wish to reduce the dimension of $X_t$ with the use of factors as a way of summarising all the available information.  Let therefore $\lbrace y_{t}, V_{t}, F_{t} \rbrace_{t=1}^T$ be a random sample, where $y_t$ is a scalar dependent variable (in this work the CPI inflation rate or the GDP growth rate), $V_t=\lbrace (V_{1,t}, V_{2,t},...) \rbrace$ is a set of targeted macroeconomic variables of countably finite dimension and $F_t=\lbrace (F_{1,t}, F_{2,t},...) \rbrace$ is a set of latent factors of countably finite dimension that need to be estimated a priori from a large panel dataset.\footnote{The vector $F_{t}$ is unobservable, therefore, in practice, we replace the infeasible vector $F_{t-1}$ with the feasible vector $\hat{F}'_{t} $, where $\hat{F}_{t}=(\hat{F}_{1,t},...,\hat{F}_{k,t}) \in \Re^k, k\in \aleph$, is the vector of  estimated factors from the panel data $X_{i,t}$.  However, given that we are not interested in coefficient inference, it is sufficient that factor estimation error has already been proven to be negligible.} \\

Without loss of generality, in the remainder of the section we assume that $Y_{1,t}=1$. Under the assumption that the conditional distribution of $y_{t+1}$ given $Z_{t}$, where $Z_{t}=(Y_{t}, V_{t}, F_{t})$,  is continuous, we can then define the $\tau^{th}$ conditional quantile of $y_{t+1}$ given $Z_{t}$ as the measurable function $q_{\tau}$ satisfying the conditional restriction
\begin{align}
P(y_{t+1} \leq q_{\tau}(Z_{t}) \mid Z_{t})=\tau, \text{ almost surely.}\quad \label{Conditional Quantile}
\end{align}\\
We consider a sequence of approximating models $m=1,2,...,M$, where the $m^{th}$ model uses $r_m$ regressors belonging to $Z_{t}$ and $M$ may go to infinity with the sample size. We write the $m^{th}$ approximating model as the Quantile Autoregression (QAR) of order p,
\begin{align}
y_{t+1} = \theta_{(m)}'Z_{t(m)}+\epsilon_{t+1}= \sum_{j=1}^{r_m} \theta_{j(m)}Z_{tj(m)} +\epsilon_{t+1}, \label{inflation}
\end{align}\\
where $ \theta_{(m)}\equiv(\theta_{1(m)},...,\theta_{r_m(m)})'$, $Z_{t(m)}=(Z_{t1(m)},...,Z_{tr_m(m)} )'$, $Z_{tj(m)}$  for $j=1,...,r_m$, are variables in $Z_t$ that appear as regressors in the $m^{th}$ model and $\theta_{j(m)}$ are the corresponding coefficients. The $\tau^{th}$ Quantile Autoregression Estimator (QARE) of $\theta_{(m)}$, proposed by \citet{koenker_quantile_2006}, is defined as 
\begin{align}
\hat{\theta}_{(m)}(\tau) &= \arg \min_{\theta_{(m)}} Q_{T(m)} (\theta_{(m)}) \\
&=\arg\min_{\theta \in \mathbb{R}^{p+1}} \sum_{t=0}^{T} \rho_{\tau} (y_{t+1}-\theta'_{(m)}Z_{t(m)}), \label{Estimator}
\end{align} \\
where $\rho_{\tau}(e)=e(\tau-\mathbbm{1}(e<0))$ is the \enquote{tick} loss function. Let $\hat{\epsilon}_{t+1(m)}(\tau)\equiv y_{t+1}-\hat{\theta}'_{(m)}(\tau)Z_{t(m)}$ be the quantile residual and let $w(\tau)\equiv(w_1(\tau),...,w_m(\tau))'$ be the weight vector in the unit simplex of $\Re^M$ and $\mathcal{W}= \lbrace w \in  [0,1]^M: \sum_{m=1}^M w_m(\tau)=1 \rbrace$. Therefore, for $t=1,...,T$, the model averaging estimator for the $\tau^{th}$ quantile is given by\\
\begin{align}
\hat{Q}_{y_{t+1}(\tau)}(w)= \sum_{m=1}^M w_{m}(\tau) Z'_{t(m)} \hat{\theta}_{(m)}(\tau). \label{ModelAverage}
\end{align}\\
It is evident in equation~(\ref{ModelAverage}) that the weight vector $w_m(\tau)$ differs across different quantiles. This implies that a heavier importance could be placed on different competing models depending on the quantile under consideration. For notational simplicity however, we drop this dependence hereinafter. Furthermore, the weight vector is independent of time. \citet{kascha_combining_2010} have previously found that time-varying weights for combining inflation density forecasts provide no advantage over fixed weights. Furthermore, in context, the computational cost of recursive weights with an extensive pool of candidate quantile models can become particularly high. As a result, unless there is strong evidence of a structural change that would imply different suitable models at each period, there is no significant gain in employing time-varying weights.\\

\vspace{0.25cm}

\subsubsection{Akaike and Bayesian Information Criteria Weights}
Both the AIC and BIC are often referred to as \enquote{smoothed} averaging \citep{buckland_model_1997}, which use exponential weights of the form $\frac{exp(\frac{-Inf_m}{2})}{\sum_{j=1}^M exp(-Inf_j)}$, where $Inf_m$ is an information criterion for the $m^{th}$ model. The aforementioned criteria assess each model fit, while at the same time penalizing for the number of estimated parameters, albeit the BIC penalises model complexity more heavily.  Both the AIC and BIC criteria are easy to compute regardless of the number of competing models $M$ we are considering and they have been proved to outperform the simplest model combination, i.e. equal weights. In the quantile regression context, following \citet{machado_robust_1993}, for the $m^{th}$ model, the AIC and BIC are respectively defined as,
\begin{align*}
AIC_m&=2ln[\frac{1}{T}\sum_{t=0}^T\rho_{\tau}(y_{t+1}-\hat{\theta}'_{(m)}(\tau)IZ{t(m)}]+2r_{m} , \quad \text{and} \\
BIC_m&=2ln[\frac{1}{T}\sum_{t=0}^T\rho_{\tau}(y_{t+1}-\hat{\theta}'_{(m)}(\tau)Z_{t(m)}]+r_{m}ln(T) ,
\end{align*}\\
where $r_m$ is the dimension of the independent vector $Z_{t(m)}$.\\

Thr AIC and BIC weights for model $m$ are thus respectively defined as,
\begin{align}
\hat{w}_{m}^{AIC}=\frac{exp(\frac{-AIC_m}{2})}{\sum_{j=1}^M exp(\frac{-AIC_j}{2})} \ \quad \text{and} \quad \hat{w}_{m}^{BIC}=\frac{exp(\frac{-BIC_m}{2})}{\sum_{j=1}^M exp(\frac{-BIC_j}{2})} .\label{AIC & BIC weights}
\end{align}

\vspace{0.25cm}

\subsubsection{Quantile Regression Information Criterion Weights}
The QRIC is a Mallows-type information criterion for QR model averaging, whose criterion function has been outlined in \citet{lu_jackknife_2015}. Mallows' type criteria tend to compare the predictive ability of subset models to that of a full model, but they still balance the trade off between obtaining a \enquote{good} model that contains as few variables as possible. Letting therefore $\hat{\epsilon}_{t+1(m)}(\tau)\equiv y_{t+1}-I_{t(m)}'\hat{\theta}_{(m)}(\tau)$ and $\hat{\epsilon}_{t+1}(w)\equiv \sum_{m=1}^M w_m \hat{\epsilon}_{t+1 (m)}(\tau)$, the QRIC can be defined as,
 \begin{align}
QRIC_T(w)&=T*Q_T(W)+\frac{\tau(1-\tau)}{f(F^{-1}(\tau))} \sum_{m=1}^M w_m r_m ,\label{QRIC}
\end{align}
where $Q_T(w)=\frac{1}{T}\sum_{t=0}^T\rho_{\tau}(\hat{\epsilon}_{t+1}(w))$ indicates the average in-sample QR prediction error. $F$ and $f$ denote the CDF and PDF, respectively, of $\epsilon_{t+1}(\tau)$ and 
$\sum_{m=1}^M w_m r_m$ signifies the number of effective parameters in the combined estimator. Therefore, in order to choose the weight vector $w$, by the QRIC,  we must estimate the sparsity function of $\epsilon_{t+1}(\tau)$, $s(\tau)=\frac{\tau(1-\tau)}{f(F^{-1}(\tau))}$. Following \citet{koenker_quantile_2005} this can be estimated by,
\begin{align}
\hat{s}(\tau)=\frac{\tilde{F}^{-1}_T (\tau+h-T)-\tilde{F}^{-1}_T (\tau-h-T)}{2*h_T}, \label{Sparsity function}
\end{align}\\
where $\tilde{F}^{-1}_T$ is an estimate of the quantile function $F^{-1}$ of $\epsilon_{t+1}(\tau)$ based on the quantile residuals obtained from the largest approximating model, $h_T=T^{-\frac{1}{5}} \lbrace  4.5 \phi^4(\Phi^{-1}(\tau))/[2\Phi^{-1}(\tau)^2+1]^2 \rbrace^{\frac{1}{5}}$ and $\phi$ and $\Phi$ are the standard normal PDF and CDF, respectively. The resulting empirical QRIC weighting vector is then defined as,\footnote{There is no closed form solution for equation~(\ref{QRIC weights}) but the optimal weight vector can be found by linear programming as in typical quantile regressions.} 
\begin{align}
\hat{w}^{QRIC} &\equiv (\hat{w}^{QRIC}_1,...,\hat{w}^{QRIC}_M)=\arg \min_{w \in \mathcal{W}} \hat{QRIC}_T(w) \notag \\
&= arg \min_{w \in \mathcal{W}} \quad [Q_T(w)+\tau(1-\tau)\hat{s}(\tau) \sum_{m=1}^M w_m r_m]. \label{QRIC weights}
\end{align}\\
It is worth noting that due to the presence of the sparsity function, the QRIC may not perform as well on extreme quantiles where the sparsity is low. However, the same holds true for the quantile regression estimator, which is why our analysis excludes the most extreme tails of the distribution and only focuses on values where $\tau \in [0.1,0.9]$.\\

\vspace{0.25cm}

\subsubsection{Jackknife Weights}
The Jackknife selection of the weighting vector $w$ is the most distinct, as rather than imposing the weighting on the fitted value of the dependent variable it, in practice, weighs the estimated quantile regression coefficients and uses the average coefficient estimator in the forecasting exercise. The Jackknife weight vector is optimal in terms of minimising the final prediction error (FPE), one of our forecast performance measures, in the sense of \citet{akaike_statistical_1970}. For all the competing models under consideration $m=1,...,M$, let $\hat{\theta}_{t(m)}$ denote the jackknife estimator of $\hat{\theta}_{(m)}$ in model $m$ with the $t^{th}$ observation excluded from the estimation. The jackknife choice of the weight vector $\hat{w}^{Jknife} =(\hat{w}^{Jknife}_1,...,\hat{w}^{Jknife}_M)$ is obtained by choosing $w \in \mathcal{W}$ to minimise the leave-one-out cross-validationn criterion function,
\begin{align}
CV_T(w)= \frac{1}{T} \sum_{t=0}^T \rho_{\tau}(y_{t+1}- \sum_{m=1}^M w_m Z'_{t(m)}\hat{\theta}_{t(m)}). \label{CV criterion function}
\end{align}\\
The resulting Jackknife weighting vector is therefore defined as,
\begin{align}
\hat{w}^{Jknife} =(\hat{w}^{Jknife}_1,...,\hat{w}^{Jknife}_M)= \arg \min_{w \in \mathcal{W}} CV_t(w).\label{Jackknife weights}
\end{align}\\
Is worth noting that, though the leave-one-out cross validaion criterion function is convex in $w$ and can be minimised by running the quantile regression of $y_{t+1}$ on $Z'_{t(m)}\hat{\theta}_{(m)}$, it cannot guarantee that the resulting solution lies in $\mathcal{W}$. However, one can express the constrained minimisation problem in (\ref{CV criterion function}) as a linear programming problem (see e.g. \citet{lu_jackknife_2015}, p. 43).\\

\vspace{0.5cm}

\subsection{Forecasting Performance Measures}
The quantile autoregression framework allows us to obtain forecasts for the conditional quantiles of the dependent variable, however, once we obtain the realisation for the dependent variable, it only provides us with the actual value of the dependent variable, but not its quantile at that period. For example, in our empirical context, we can forecast the $\tau^{th}$ quantile of the inflation rate $H$-periods ahead, however, fast forward $H$ periods later what we obtain is the value of the inflation rate, $y_{t+H}$, and not the value of the $\tau^{th}$ quantile, $Q_{y_{t+H}}(\tau)$. This implies that evaluating the forecast performance of the aforementioned weighting methodologies is not a trivial issue, but it also allows for a certain degree of flexibility.  We therefore evaluate the suggested methodologies using two different forecast performance measures.\\

\vspace{0.25cm}

\subsubsection{Unconditional Coverage}
Though we may not be able to obtain an observation for the realisation of the $\tau^{th}$ quantile, in practice quantile forecasts correspond to one-sided interval forecasts of the dependent variable of interest. Therefore, the unconditional coverage testing framework of \citet{christoffersen_evaluating_1998} poses a suitable measure for evaluating the \enquote{accuracy} of the quantile forecast.  Correct coverage tests in practice aim at testing whether a sequence of conditional quantile forecasts satisfies certain optimality conditions. Here, we are not interested in testing in absolute terms if the quantile forecasts are correct, but which method for choosing the weight vector might be better, thus we are interested into which method provides as with the most correct coverage rate. \\

Assume we obtain a sequence of out-of-sample forecasts,$\lbrace \hat{q}_{t+h|t}(\tau) \rbrace_{t=0}^P$, of the conditional quantile $Q_{y_{t+H}}(\tau)$ of the time series $y_{t+H}$. This quantile forecast is an upper limit for an interval forecast  for the time series $y$, for time $t+H$, made at time $t$, for the coverage probability $\tau$.\footnote{If the $\tau^{th}$ quantile forecast, $q_{t+H|t}(\tau)$ is accurate, then in practice the number of times that the realisation of $y_{t+H}$ falls below the forecast value should on average equal the nominal quantile level for which we are forecasting.}  However, as we move across the quantiles and towards the upper tail of the distribution, it is more likely that most of the observations will fall below the estimated quantile, which can result in better coverage results by construction, rather than due to better forecast performance. Therefore, for each quantile under evaluation, we choose to obtain a two-sided interval forecast, $\lbrace (L^{\tau}_{t+H|t}(p), U^{\tau}_{t+H|t}(p)) \rbrace_{t=0}^P$, where $L^{\tau}_{t+H|t}(p)$ and $U^{\tau}_{t+H|t}(p)$ are the lower and upper limits of the ex-ante interval forecast for quantile $\tau$, for time $t+H$, made at time t, for the nominal coverage probability, p. We can then define an indicator variable, $\mathcal{I}^{\tau}_{t+H}$ for time $t+H$,  where
\begin{align}
\mathcal{I}^{\tau}_{t+H}= \left\lbrace
\begin{array} {ll}
1, \quad  &if \quad y_{t+H} \in [L^{\tau}_{t+H|t}(p), U^{\tau}_{t+H|t}(p)]\\
0, \quad  &if \quad y_{t+H} \notin  [L^{\tau}_{t+H|t}(p), U^{\tau}_{t+H|t}(p)]
\end{array}
\right. \label{Indicator Function}
\end{align}\\

We can therefore measure the efficiency of this forecast interval by measuring the amount of times that the indicator variable takes a value of 1 as a proportion of the total number of prediction periods. For example, in the one step ahead forecast, the coverage rate is defined as
\begin{align}
\text{Coverage rate(w)}= \frac{1}{P} \sum_{s=0}^{P} \mathcal{I}^{\tau}_{s+1} ,\label{Coverage}
\end{align}\\
where $P$ s the number of periods from the sample set aside for forecast evaluation. The quantile forecast is dependent on the weight vector chosen by each method, though this dependence has been dropped for simplicity, therefore, the closer the empirical coverage rate to the nominal coverage probability, p, the more \enquote{accurate} the method in terms of approximating the true quantile of the distribution.\\

Although the unconditional coverage measure is a suitable evaluation method for quantile forecasts, it can only provides us with an indication of whether the realised observation falls within the forecast interval or not. In case the realised observation falls outside the interval forecast, implying we have a violation, we have no information on how far away it is. It is true that in the spirit of quantile regression the only relevant issue is only whether or not the observation falls below an estimated quantile and no attention is given on how far an observation is from the quantile. Nevertheless, in this empirical context where the prediction period is rather small and thus empirical coverage rates might not provide a clear image on which methodology performs better, we also wish to examine the behaviour of observations that fall outside the interval forecast. We therefore complement the unconditional coverage measure with the interval score of \citet{gneiting_strictly_2007}. This scoring rule has an intuitive appeal in that the forecaster incurs a penalty, the size of which depends on the relevant confidence level, if the observation misses the interval, but is also rewarded for narrow prediction intervals. In our empirical context, the interval score can be defined for a nominal coverage probability $p$ as,
\begin{align}
S^{\tau}(L,U; y_{t+h})&=(U-L)+\frac{2}{1-p}(L-y_{t+h})\mathbbm{1}(y_{t+h}<L)+\frac{2}{1-p}(y_{t+h}-U)\mathbbm{1}(y_{t+h}>U),
\end{align}\\
where the dependencies of the lower and upper bound, $L$ and $U$, on $\tau$, $t$, $t+h$ and $p$ has been dropped for simplicity. It is evident that this scoring rule imposes the same penalty for violations that occur below the lower bound and above the upper bound. It is worth mentioning that one could examine an asymmetric form of penalisation that would take into consideration the spectrum of the distribution available below the lower bound or above the upper bound, which is dependent on the quantile of interest considered, since this could impact the possible magnitude of a violation. However, in this context, this is not expected to influence the results in a significant way and has not been considered.\\

We therefore obtain the one-step-ahead average interval score as,
\begin{align}
\text{Interval Score}(w)= \frac{1}{P} \sum_{s=0}^{P} S^{\tau}(L^{\tau}_{s+1|s}(p),U^{\tau}_{s+1|s}(p); y_{s+1}), \label{Interval Score}
\end{align}\\
where $P$ s the number of forecast evaluation periods and the dependence of the Interval Score on the weight vector $w$ is due to the dependence of lower and upper bounds of our confidence intervals on $w$, which has been dropped for simplicity. In this case a lower interval score is desirable, since it implies that even if the realised observation falls outside the forecast interval, the magnitude of the violation is not large.\\

\vspace{0.25cm}

\subsubsection{Final Prediction Error}
Different loss functions $\mathcal{L}$ arguably correspond to different optimal forecasts. For example, letting $\hat{\epsilon}_{t+H}\equiv y_{t+H}-\hat{q}_{t+H|t}$ be the forecast error, if a quadratic loss function is used such that $\mathcal{L}(\hat{\epsilon}_{t+H})=\hat{\epsilon}_{t+H}^2$, then the optimal forecast for $y_{t+H}$ would be conditional mean. Similarly, if the absolute value loss function is used such that $\mathcal{L}(\hat{\epsilon}_{t+H})=|\hat{\epsilon}_{t+H}|$, then the optimal forecast for $y_{t+H}$ would correspond to the conditional median. Based on this idea, \citet{giacomini_evaluation_2005} argue that, in the case conditional quantiles, the corresponding loss function would be the asymmetric linear loss function, $\mathcal{L}(\hat{\epsilon}_{t+H})\equiv \rho_{\tau}(\hat{\epsilon}_{t+H})\equiv \hat{\epsilon}_{t+H}(\tau-\mathbbm{1}(\hat{\epsilon}_{t+H}<0))$ and therefore the optimal forecast for $y_{t+H}$ is its conditional $\tau^{th}$ quantile. Therefore, in the model averaging framework, an alternative forecast performance measure for the one-step-ahead forecast is the Final Prediction Error (FPE), which, as a function of the weight vector, is defined as, 
\begin{align}
FPE_P(w)= \frac{1}{P} \sum_{s=0}^P [\rho_{\tau} (y_{s+1}-\sum_{m=1}^M \hat{w}_m\hat{\theta}'_{(m)}Z_{s(m)})], \label{FPE}
\end{align}\\
where $\hat{w}_m$ is chosen by one of the suggested methodologies.  In this case, the parameter $\tau$ describes the degree of asymmetry in the loss function, where a value less than one-half indicates that over-predicting results to a greater loss for the forecaster than under-predicting by the same magnitude.\footnote{When the parameter $\tau$ equals one-half,  over-predicting and under-predicting generates the same loss, thus we converge to the absolute value loss function where the optimal forecast is the conditional median.} Therefore, the smaller the FPE, the better the weight vector method in terms of the out-of-sample quantile prediction error. \\

\vspace{0.5cm}

\section{Forecasting the Distribution of GDP Growth and CPI Inflation}
Inflation is one of the most important variables due to its dominant role in many macroeconomic models (\citealp{levin_is_2002, angeloni_new_2006}). Quarterly \enquote{Inflation Reports} have become the new norm for the majority of central banks and even though the costs and benefits of transparency are still widely debated, it is broadly agreed that a central bank should be concerned with inflation forecasting. Nevertheless, as argued by, inter alia, \citet{faust_forecasting_2013, henry_is_2004}, a point forecast of inflation without some measure of associated uncertainty is arguably of little value. Policymakers forecasting inflation need to not only consider the most likely outcome for inflation, but all possible paths that inflation can take, which involves examining the dynamics and higher moments of inflation. Similarly, over the recent years policy-makers have shifted focus towards downside risk for GDP growth rather than point forecasts for the conditional mean of growth. This is due to the fact that such point forecasts ignore the risks surrounding these central forecasts and thus may paint an overly optimistic picture of the state of the economy \citep{adrian_vulnerable_2019}. A density forecast therefore gives a more complete characterisation of future growth and inflation prospects and forecasting future conditional quantiles is a computationally easy method to obtain it, while allowing different quantiles to exhibit different sensitivity to predictors.\\

\vspace{0.25cm}

\subsection{Data}
In this empirical study we examine, under a model averaging approach, which methodology for choosing the appropriate weight vector is better suited for forecasting the one-quarter-ahead annual GDP growth rate and CPI inflation rate. Several of our competing models will involve latent factors as a way to summarise a large amount of information from different macroeconomic variables. We therefore consider for the estimation of factors series containing data on inflation, real activity and indicators of money and key asset prices for the United Kingdom. We will undertake the analysis using a sample which includes quarterly data from 174 macroeconomic variables spanning from the second quarter of 1991 to the second quarter of 2018, with a total of $T=109$ observations.\footnote{This dataset is a subset of the dataset used by \citet{ellis_what_2014} to create a time-varying factor augmented VAR model for the UK monetary transmission mechanism and details regarding the variable used can be found in the Appendix.} All the data has been stationarised prior to use.  Given that latent factors have been proven important in modelling and predicting our variables of interest we will extract two latent factors from the macroeconomic dataset.
The latent factors are extracted using a recursive estimation scheme, thus in each period within the prediction sample all available past information is utilised, but the results hold under alternative estimation schemes as well. Furthermore, given the close relationship between growth and inflation, we shall include these variables as individual regressors. Therefore, there would overall be 5 possible regressors for each dependent variable, as shown in Tables~\ref{table:Regressors Growth} $\&$~\ref{table:Regressors Inflation}.\\

\begin{table}[H]
\center
\begin{tabular}{lllr}
\hline
\hline
Regressor & Name                             &  & \multicolumn{1}{c}{Correlation with GDP growth} \\ \hline
$r_1$         & First lag of GDP Growth  &  & 0.9048                               \\
$r_2$          & Second lag of GDP Growth &  & 0.7020                                \\
$r_3$          & Lagged CPI Inflation    &  & -0.4453                                \\
$r_4$          & Lagged first latent factor       &  & 0.2354                                \\
$r_5$          & Lagged second latent factor      &  &  0.2584                           \\ \hline \hline
\end{tabular}
\caption{Regressors for the one-quarter-ahead GDP growth rate} 
\label{table:Regressors Growth}
\end{table}

\vspace{0.5cm}

\begin{table}[H]
\center
\begin{tabular}{lllr}
\hline
\hline
Regressor & Name                             &  & \multicolumn{1}{c}{Correlation with CPI inflation} \\ \hline
$r_1$          & First lag of CPI inflation  &  & 0.9179                                \\
$r_2$          & Second lag of CPI inflation &  & 0.8034                                \\
$r_3$          & Lagged GDP Growth   &  & -0.3667                                \\
$r_4$          & Lagged first latent factor       &  & 0.1274                                \\
$r_5$          & Lagged second latent factor      &  & -0.0793                           \\ \hline \hline
\end{tabular}
\caption{Regressors for the one-quarter-ahead CPI inflation rate} 
\label{table:Regressors Inflation}
\end{table}

\vspace{0.25cm}

\subsection{Competing Models}
For each dependent variable of interest we have constructed 57 non-nested candidate models with all the possible combinations between the five regressors and an intercept, where the smallest models has at least two regressors. Our smallest and largest model can therefore be characterised by the following regressors, $\lbrace 1, r_1 \rbrace$ and $\lbrace 1,r_1, r_2, r_3, r_4, r_5 \rbrace$, respectively. We split the sample into an estimation sample of size $T_1=\frac{T}{2}$, used to determine the appropriate weight vector for model averaging and an evaluation sample of size $T_2=T-T_1$, used for forecast performance evaluation. We then choose the relevant weight vector for the 57 competing models under the four methodologies presented and then using the averaged model construct one-period-ahead forecasts for 9 different quantiles where $\tau \in [0.1,0.9]$. We also construct forecasts with the naive quantile autoregressive model of order 1, a full model which includes all the aforementioned regressors, as well as an average model which assigns an equal weight to all 57 competing models.  \\

\vspace{0.25cm}

\subsection{Out-of-Sample Performance}
As it was outlined in the framework section, the 57 competing models get assigned a corresponding weight for model averaging by each of the four methodologies, which are then used to obtain a forecast average of the one-step-ahead GDP growth and CPI inflation rate. Tables~\ref{table:Growth Weights AIC} -\ref{table:Inflation Weights Jackknife} in the Appendix demonstrate the allocated weights by each methodology across the 57 competing models, for each of the two dependent variables. It is evident that the AIC and BIC allocate similar weights, which are roughly equal across  all the models under consideration, so their out-of-sample performance is expected to be similar. On the other hand, the QRIC allocates significantly high weights to specific models, assigning a zero weight to the majority of the competing models. It is also evident that this criterion favours low-dimensional models, in contrast with the Jackknife method, which favours higher dimension models. Furthermore, Tables~\ref{table:GDP Coverage} -\ref{table:CPI FPE} in the Appendix show the empirical coverage rates and final prediction error for all the methodologies under consideration. From those results, as expected given the allocated weights, the out-of-sample performance of a simple average model of equal weights is comparable to the performance of an average model where weights are assigned using the AIC or BIC, for both growth and inflation. For brevity purposes therefore, in the remainder of the paper we shall be comparing the forecasting performance of the naive model, the full model and the average models with an equal weighting and with weighting assigned by the QRIC and Jackknife criterion. \\

\begin{figure}[H]
\center
\includegraphics[scale=0.085]{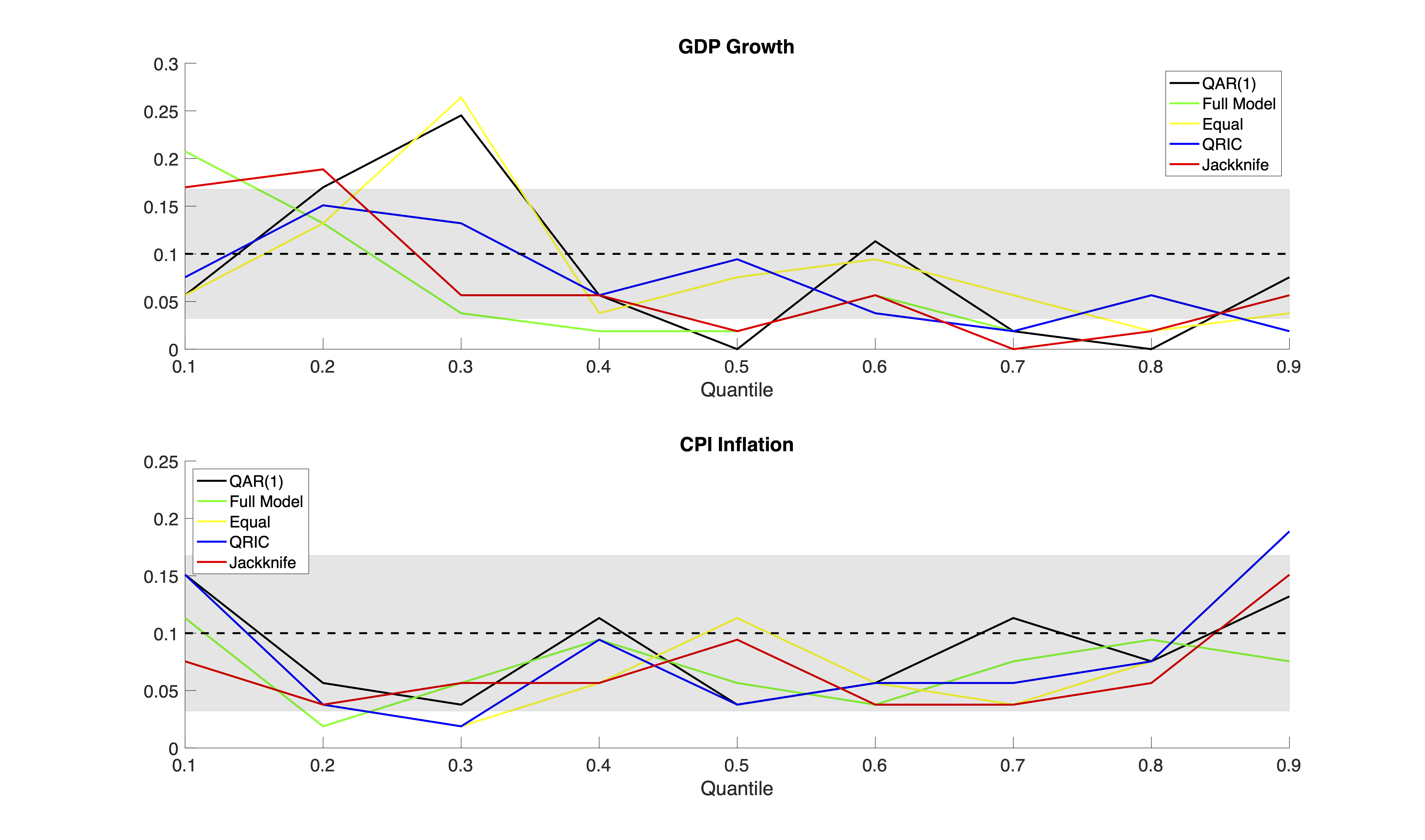}
\caption{Out-of-Sample Coverage Rates across quantiles with $10\%$ confidence bands}
\label{figure: Coverage}
\end{figure}

Figure~\ref{figure: Coverage} shows the empirical coverage rate by each of the remaining competing methods. The top panel has GDP growth as the dependent variable and the bottom panel is for CPI inflation. We evaluate $9$ equidistributed points for $\tau \in [0.1, 0.9]$, where for each quantile of interest we obtain an interval forecast with a nominal coverage probability of $10\%$.\footnote{For example, if the quantile of interest is the conditional median, we obtain an interval forecast by estimating the $45^{th}$ and $55^{th}$ conditional quantile} The shaded region demonstrates the confidence interval for the null hypothesis that the empirical coverage rate is equal to nominal coverage probability of $10\%$.\footnote{With larger nominal coverage probabilities (e.g. $20\%$), the absolute performance of the methodologies, in particular with respect to coverage, improves due to the larger forecast intervals, but the comparative performance between methodologies remains identical.}\\

Focusing on GDP, in terms of coverage, although it is not clear whether a single methodology outperforms all its competitors, several conclusions can be made. First, we can see that all the methodologies tend to provide more conservative interval forecasts at the lower tails of the distribution, but such intervals become more liberal as we move towards the upper tail. Furthermore, on several occasions we see that the null hypothesis of nominal coverage of $10\%$ is not satisfied at certain points of the distribution by several methodologies, indicating that perhaps additional regressors should be considered. Nevertheless, overall, the QRIC weighting methodology seems to have a rather competitive performance, with empirical coverage rates close to the nominal probability for a substantial range of the distribution. More interestingly, the fact that the full model does not outperform several of the model averaging methodologies proves that there are gains to be made by considering an array of competing models, even if such models do not utilise the full information. This might be, as argued by \citet{granger_can_1996}, due to presence of different estimation biases in the competing models that may cancel each other out and, as a result, provide us with a more accurate forecast than the full model.\\

With respect to CPI inflation, there exists a much clearer picture. Firstly, all the methodologies provide liberal forecast intervals for the majority of the distribution, with the exception of the most extreme tails. A simple average is often outperformed by the QRIC and Jackknife methodologies, but overall it is evident that the naive QAR(1) model seems to be outperforming all the competitors, by achieving coverage rates that are the closest to $10\%$ for the majority of the quantiles of interest.\\

\begin{figure}[H]
\center
\includegraphics[scale=0.085]{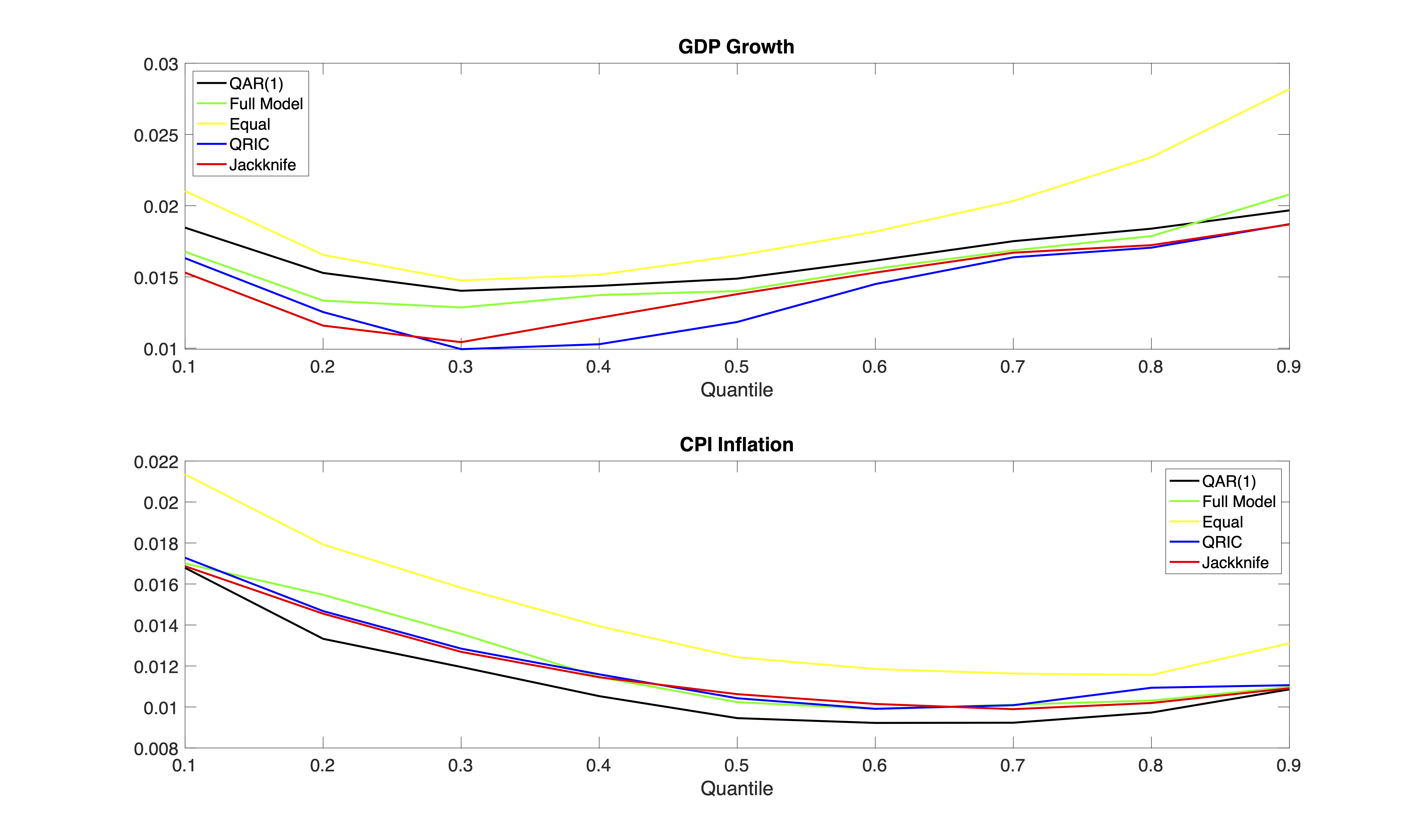}
\caption{Out-of-Sample Forecast Error Loss across quantiles}
\label{figure: Interval Score}
\end{figure}
\bigskip

\vspace{0.5cm}

However, taking into consideration the interval score can provide more clarity to our conclusions.\footnote{It is worth noting once again that one could consider an asymmetric penalty, in order to penalise violations according to which quantile of the distribution once is interested and the maximum possible violation that can incur at that point.} As it is evident in Figure~\ref{figure: Interval Score}, the interval score for the QRIC and Jackknife approaches is significantly lower when compared to the other methodologies, particularly in the case of GDP growth. This result indicates that even though the QRIC and Jackknife approaches may have a significant number of violations (i.e. the realised observation falls outside the interval forecast) and thus not as accurate coverage rates, the magnitude of the violation is not large and/or the prediction intervals provided by these methodologies are narrower. This implies that for several of the prediction periods, the realised observation is not far from the lower or upper bound of the forecast interval. In contrast, the simple average model may have a similar number of violations, but when a violation takes place the forecast error is significantly larger. Furthermore, similarly to what we have identified before, the full model seems to be outperformed by certain model averaging methodologies.\\

When forecasting CPI inflation, a similar picture is painted as the one for GDP growth, in that a simple average is associated with larger mean forecast errors across the whole distribution. Similar with its coverage performance, the QAR(1) seems to be outperforming all the methodologies under consideration, implying that the naive model might be the best for predicting the quantiles of CPI inflation. Another take away from this measure of CPI inflation is the fact that the interval score and thus the forecast error across all the methodologies seems to be smaller in the upper tail of the distribution. This could be explained by a higher variation in the right tail of the CPI inflation distribution, as this has been found in \citet{phella_consistent_2020}, which enables for better forecasting regardless of the allocation of weights across the different competing models. Conversely, downturns are more difficult to predict, probably due to the zero lower bound that is present, thus lower quantiles are associated with larger violations. \\

Looking in conjunction at the two figures, we can conclude that, in the case of GDP growth, the QRIC and, to a certain extent, the Jackknife model averages can produce reliable forecasts for a large spectrum of the distribution, though such forecasts tend to be rather liberal. Notably however, for CPI inflation, the QAR(1) seems to be performing the best with coverage rates close to $10\%$ across most of the evaluated quantiles and with violation magnitudes lower than the ones produced by the QRIC and Jackknife. This is in line with the notion present in the relevant literature that past inflation is the best predictor for future inflation.  On the other hand, the fact that the QRIC and Jackknife model averages for GDP growth outperform the full model,  demonstrates the importance of considering the aggregation of multiple competing models. Furthermore, the fact that for GDP growth the methodologies that forecast better are the ones which attribute significant weights to models with latent factors (see Table~\ref{table:Growth Weights QRIC} in Appendix), is an indication that such latent factors are relevant for out-of-sample estimation of conditional quantiles of growth. This is a complementary result to that present in the in-sample estimation literature, where latent factors, summarising a larger information set, were found to carry relevant information for in sample estimation of GDP growth, but not CPI inflation.\\

\begin{figure}[H]
\center
\includegraphics[scale=0.085]{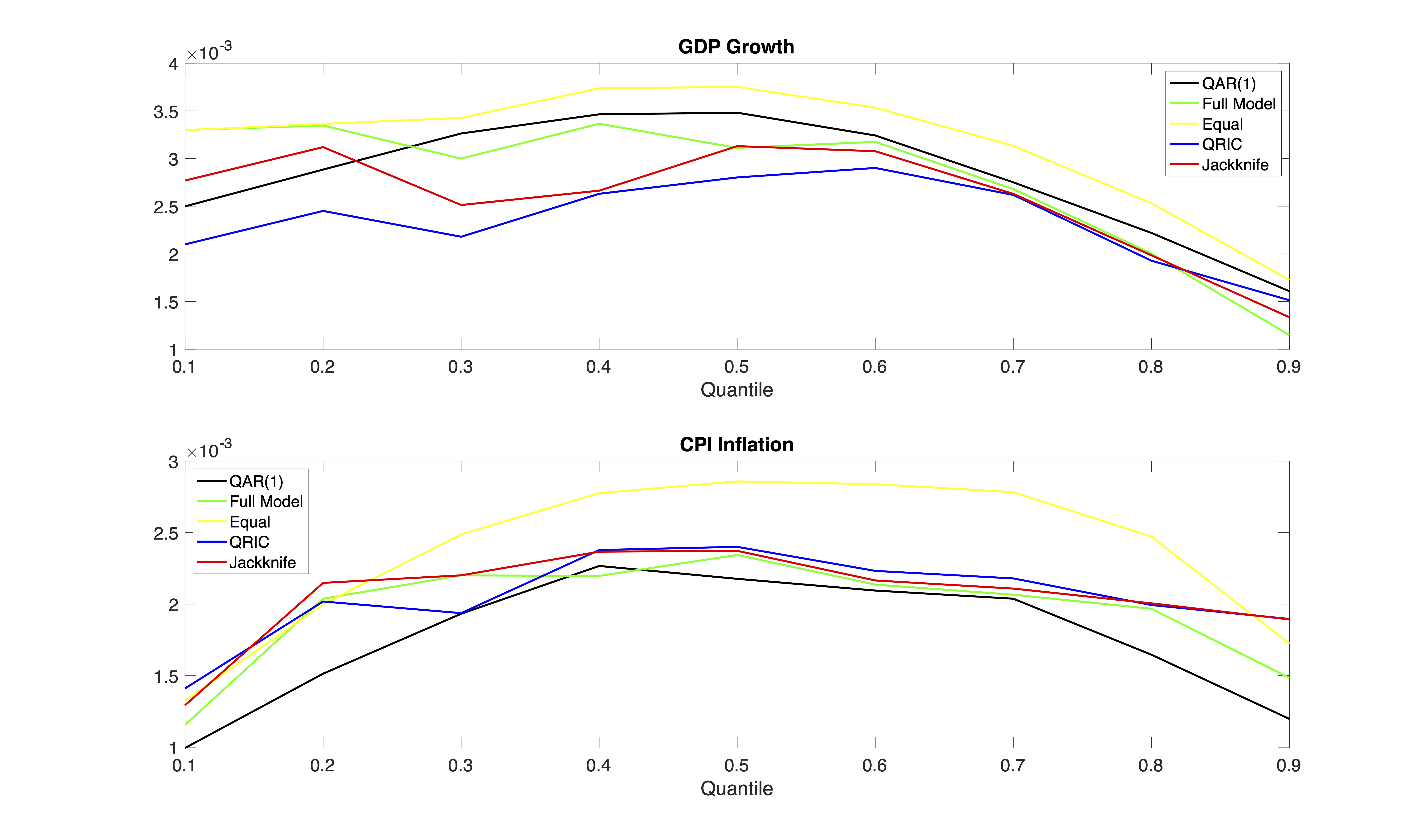}
\caption{Out-of-Sample Final Prediction Error across quantiles}
\label{figure: FPE}
\end{figure}
 
\vspace{0.5cm}

The evidence when considering the final prediction error are much clearer than that for coverage. It is clear in both panels of  Figure~\ref{figure: FPE} that the final prediction error of the QRIC and Jackknife methodologies seem to be the best for most of the quantiles considered, albeit the full model is competitive in the extreme right tail.  This implies that a model average with distinctive weighting serves as a better predictor for $y_{t+1}$. The good performance of the QRIC methodology is not surprising in this case, as the quantile regression criterion was constructed in order to minimise this type of prediction error.  Once again, we see that the naive QAR(1) model of CPI inflation outperforms all the competing methodologies, implying that conditional quantiles estimated solely with past inflation information can act as the best predictor for the future value of inflation. This is a striking difference with mean regressions of the UK CPI inflation, where forecast performance relative to the AR benchmark model is improved when forecasts are combined \citep{kapetanios_forecast_2008}, though this work utilises information from multiple data sources and data types in the forecast combination, including subjective data. \\

The distinctive differences and similarities between the two performance measures in all three figures highlights certain empirical facts. Firstly, it is evident that despite the computational ease,  assigning weights according to the Akaike and Bayesian Information Criteria does not improve forecasting performance over a model averaging approach where each competing model is assigned an equal weight. Most importantly, the distinctive performance of the different model averaging methodologies demonstrates the importance of choosing the assigned weights correctly. Secondly, the differences between the forecasting performance of the naive QAR(1) models in the case of GDP growth versus CPI inflation, demonstrates fundamental differences regarding which variables are the best predictors in each individual case, despite the close relationship between inflation and growth.\\

Similarly with the results of in-sample estimation found in \citet{phella_consistent_2020}, past inflation is the best predictor for the future path of inflation, while GDP growth can be predicted best if additional latent factors and models which include such factors are taken into consideration.  The former has been long discussed in the case of mean regressions of inflation, however these results prove that this remains true in the case of conditional quantiles as a way to obtain a trace of the inflation distribution and therefore a measure of the risk and uncertainty surrounding future inflation. Lastly, the good performance in forecasting GDP growth of the QRIC and Jackknife approach, which assign higher weights in models with latent factors, is also in line with the in-sample literature, where these latent factors had been shown to have a distinctive impact across different parts of the growth distribution. These results once again highlight the necessity of high dimensional macroeconomic data when extracting latent factors and at the same time demonstrates the importance of such latent factors in quantile regression models for different variables of interest.\\

\vspace{0.5cm}

\section{Conclusion}
We have constructed forecasts of the conditional quantiles for the GDP growth rate and CPI Inflation rate of the United Kingdom, using a model averaging approach where the weight vector has been chosen by different criteria: the AIC, BIC, QRIC and Jackknife. The literature has long supported that forecast combinations of average inflation can improve out-of-sample performance, however this literature ignores the risks and uncertainties around central forecasts. Similarly for growth, although Bayesian model averages have long been used, frequentist model averaging has not gained significant attention. This work addresses both issues, by dealing with the conditional quantiles of growth and inflation and thus their whole distributions.\\

We find that, in terms of coverage the distinctive weighting across competing models imposed by the QRIC and Jackknife can outperform an equal weighting, as well as the full model, for forecasting GDP growth, when an interval score is simultaneously considered. On the other hand, the benchmark QAR(1) model of CPI inflation outperforms all model averaging methodologies. In terms of final prediction error, a similar picture exists. Overall, the results demonstrate the importance of high dimensional macroeconomic data and the inclusion of latent factors, as a way to summarise such data, when forecasting GDP growth but confirm that past inflation is the bet predictor for future inflation. Previously, latent factors were shown to be relevant for in-sample estimations of the growth distribution. Although good in-sample performance does not guarantee good out-of-sample performance, this work demonstrates that in the case of growth, latent factors are also relevant and important in out-of-sample forecasts of the conditional quantiles.

\newpage
\section{Appendix} 
\subsection{Weight Allocation}
\begin{center}
\begin{minipage}{1.00\linewidth}
  \singlespacing \centering \flushleft \noindent
\begin{table}[H]
\center
\caption{Included Regressors in Competing Models}
%\vspace{0.5cm}
\begin{threeparttable}
\scriptsize{\begin{tabular}{lllllll}
\hline
\hline
Competing Model ID & \multicolumn{6}{c}{Included Regressors} \\ \hline
\textbf{1}         & 1   & 2   & NaN   & NaN   & NaN  & NaN  \\
\textbf{2}         & 1   & 3   & NaN   & NaN   & NaN  & NaN  \\
\textbf{3}         & 1   & 4   & NaN   & NaN   & NaN  & NaN  \\
\textbf{4}         & 1   & 5   & NaN   & NaN   & NaN  & NaN  \\
\textbf{5}         & 1   & 6   & NaN   & NaN   & NaN  & NaN  \\
\textbf{6}         & 2   & 3   & NaN   & NaN   & NaN  & NaN  \\
\textbf{7}         & 2   & 4   & NaN   & NaN   & NaN  & NaN  \\
\textbf{8}         & 2   & 5   & NaN   & NaN   & NaN  & NaN  \\
\textbf{9}         & 2   & 6   & NaN   & NaN   & NaN  & NaN  \\
\textbf{10}        & 3   & 4   & NaN   & NaN   & NaN  & NaN  \\
\textbf{11}        & 3   & 5   & NaN   & NaN   & NaN  & NaN  \\
\textbf{12}        & 3   & 6   & NaN   & NaN   & NaN  & NaN  \\
\textbf{13}        & 4   & 5   & NaN   & NaN   & NaN  & NaN  \\
\textbf{14}        & 4   & 6   & NaN   & NaN   & NaN  & NaN  \\
\textbf{15}        & 5   & 6   & NaN   & NaN   & NaN  & NaN  \\
\textbf{16}        & 1   & 2   & 3     & NaN   & NaN  & NaN  \\
\textbf{17}        & 1   & 2   & 4     & NaN   & NaN  & NaN  \\
\textbf{18}        & 1   & 2   & 5     & NaN   & NaN  & NaN  \\
\textbf{19}        & 1   & 2   & 6     & NaN   & NaN  & NaN  \\
\textbf{20}        & 1   & 3   & 4     & NaN   & NaN  & NaN  \\
\textbf{21}        & 1   & 3   & 5     & NaN   & NaN  & NaN  \\
\textbf{22}        & 1   & 3   & 6     & NaN   & NaN  & NaN  \\
\textbf{23}        & 1   & 4   & 5     & NaN   & NaN  & NaN  \\
\textbf{24}        & 1   & 4   & 6     & NaN   & NaN  & NaN  \\
\textbf{25}        & 1   & 5   & 6     & NaN   & NaN  & NaN  \\
\textbf{26}        & 2   & 3   & 4     & NaN   & NaN  & NaN  \\
\textbf{27}        & 2   & 3   & 5     & NaN   & NaN  & NaN  \\
\textbf{28}        & 2   & 3   & 6     & NaN   & NaN  & NaN  \\
\textbf{29}        & 2   & 4   & 5     & NaN   & NaN  & NaN  \\
\textbf{30}        & 2   & 4   & 6     & NaN   & NaN  & NaN  \\
\textbf{31}        & 2   & 5   & 6     & NaN   & NaN  & NaN  \\
\textbf{32}        & 3   & 4   & 5     & NaN   & NaN  & NaN  \\
\textbf{33}        & 3   & 4   & 6     & NaN   & NaN  & NaN  \\
\textbf{34}        & 3   & 5   & 6     & NaN   & NaN  & NaN  \\
\textbf{35}        & 4   & 5   & 6     & NaN   & NaN  & NaN  \\
\textbf{36}        & 1   & 2   & 3     & 4     & NaN  & NaN  \\
\textbf{37}        & 1   & 2   & 3     & 5     & NaN  & NaN  \\
\textbf{38}        & 1   & 2   & 3     & 6     & NaN  & NaN  \\
\textbf{39}        & 1   & 2   & 4     & 5     & NaN  & NaN  \\
\textbf{40}        & 1   & 2   & 4     & 6     & NaN  & NaN  \\
\textbf{41}        & 1   & 2   & 5     & 6     & NaN  & NaN  \\
\textbf{42}        & 1   & 3   & 4     & 5     & NaN  & NaN  \\
\textbf{43}        & 1   & 3   & 4     & 6     & NaN  & NaN  \\
\textbf{44}        & 1   & 3   & 5     & 6     & NaN  & NaN  \\
\textbf{45}        & 1   & 4   & 5     & 6     & NaN  & NaN  \\
\textbf{46}        & 2   & 3   & 4     & 5     & NaN  & NaN  \\
\textbf{47}        & 2   & 3   & 4     & 6     & NaN  & NaN  \\
\textbf{48}        & 2   & 3   & 5     & 6     & NaN  & NaN  \\
\textbf{49}        & 2   & 4   & 5     & 6     & NaN  & NaN  \\
\textbf{50}        & 3   & 4   & 5     & 6     & NaN  & NaN  \\
\textbf{51}        & 1   & 2   & 3     & 4     & 5    & NaN  \\
\textbf{52}        & 1   & 2   & 3     & 4     & 6    & NaN  \\
\textbf{53}        & 1   & 2   & 3     & 5     & 6    & NaN  \\
\textbf{54}        & 1   & 2   & 4     & 5     & 6    & NaN  \\
\textbf{55}        & 1   & 3   & 4     & 5     & 6    & NaN  \\
\textbf{56}        & 2   & 3   & 4     & 5     & 6    & NaN  \\
\textbf{57}        & 1   & 2   & 3     & 4     & 5    & 6  \\ \hline \hline
\end{tabular}}
\begin{tablenotes}
            \footnotesize{\item $1=Intercept, 2=r_1,3=r_2, 4=r_3,5=r_4  \text{ and } 6=r_5$.}
\end{tablenotes}
\end{threeparttable}
\label{Competing Models ID}
\end{table}
\end{minipage}
\end{center}
\vspace{1cm}

\begin{table}[H]
\center
\caption{Weight allocation by AIC for quantiles of interest for GDP Growth}
\vspace{0.5cm}
\footnotesize{\begin{tabular}{p{0.1\textwidth}lllllllll}
\hline
\hline
Quantiles Model ID &0.1&0.2&0.3&0.4&0.5&0.6&0.7&0.8&0.9\\
\hline
1&0.02&0.02&0.02&0.02&0.02&0.02&0.02&0.02&0.02\\
2&0.018&0.019&0.019&0.019&0.019&0.019&0.019&0.019&0.019\\
3&0.019&0.019&0.019&0.019&0.019&0.019&0.019&0.018&0.018\\
4&0.017&0.017&0.018&0.018&0.018&0.018&0.018&0.018&0.018\\
5&0.017&0.017&0.017&0.018&0.018&0.018&0.018&0.018&0.018\\
6&0.02&0.02&0.02&0.02&0.02&0.02&0.019&0.019&0.018\\
7&0.019&0.019&0.019&0.02&0.02&0.02&0.02&0.02&0.02\\
8&0.019&0.019&0.019&0.019&0.019&0.019&0.019&0.019&0.019\\
9&0.019&0.019&0.019&0.019&0.019&0.019&0.019&0.019&0.018\\
10&0.018&0.018&0.018&0.018&0.018&0.018&0.018&0.018&0.018\\
11&0.018&0.018&0.018&0.018&0.018&0.018&0.017&0.017&0.017\\
12&0.018&0.018&0.018&0.018&0.017&0.017&0.017&0.017&0.016\\
13&0.016&0.016&0.015&0.015&0.015&0.015&0.015&0.015&0.015\\
14&0.016&0.015&0.015&0.015&0.015&0.015&0.015&0.015&0.015\\
15&0.016&0.015&0.014&0.014&0.013&0.012&0.012&0.011&0.01\\
16&0.019&0.019&0.019&0.019&0.019&0.019&0.02&0.02&0.02\\
17&0.019&0.019&0.019&0.019&0.019&0.019&0.019&0.019&0.019\\
18&0.019&0.019&0.019&0.019&0.019&0.019&0.019&0.019&0.02\\
19&0.019&0.019&0.019&0.019&0.019&0.019&0.019&0.019&0.019\\
20&0.018&0.018&0.018&0.018&0.018&0.018&0.018&0.018&0.018\\
21&0.018&0.018&0.018&0.017&0.018&0.018&0.018&0.018&0.018\\
22&0.017&0.018&0.018&0.017&0.017&0.018&0.018&0.018&0.018\\
23&0.018&0.018&0.018&0.018&0.018&0.018&0.018&0.018&0.018\\
24&0.018&0.017&0.017&0.017&0.017&0.017&0.017&0.017&0.017\\
25&0.016&0.016&0.016&0.016&0.016&0.017&0.017&0.017&0.017\\
26&0.018&0.019&0.019&0.019&0.019&0.019&0.019&0.019&0.019\\
27&0.018&0.019&0.019&0.019&0.019&0.018&0.018&0.018&0.018\\
28&0.018&0.019&0.019&0.019&0.019&0.018&0.018&0.018&0.017\\
29&0.018&0.018&0.018&0.018&0.018&0.018&0.018&0.018&0.019\\
30&0.018&0.018&0.018&0.018&0.018&0.018&0.018&0.018&0.019\\
31&0.018&0.018&0.018&0.018&0.018&0.018&0.018&0.018&0.018\\
32&0.017&0.017&0.017&0.017&0.017&0.017&0.017&0.017&0.017\\
33&0.017&0.017&0.017&0.017&0.017&0.017&0.017&0.017&0.017\\
34&0.017&0.017&0.016&0.016&0.016&0.016&0.016&0.016&0.016\\
35&0.015&0.014&0.014&0.014&0.014&0.014&0.014&0.014&0.014\\
36&0.018&0.018&0.018&0.018&0.018&0.018&0.018&0.018&0.019\\
37&0.018&0.018&0.018&0.018&0.018&0.018&0.018&0.018&0.018\\
38&0.018&0.018&0.018&0.018&0.018&0.018&0.018&0.018&0.018\\
39&0.018&0.018&0.018&0.018&0.018&0.018&0.018&0.018&0.018\\
40&0.018&0.018&0.017&0.018&0.018&0.018&0.018&0.018&0.018\\
41&0.018&0.017&0.017&0.018&0.018&0.018&0.018&0.018&0.018\\
42&0.017&0.017&0.017&0.017&0.017&0.017&0.017&0.017&0.017\\
43&0.017&0.017&0.017&0.017&0.017&0.017&0.017&0.017&0.017\\
44&0.016&0.016&0.016&0.016&0.016&0.016&0.016&0.017&0.017\\
45&0.017&0.017&0.016&0.016&0.016&0.016&0.016&0.016&0.017\\
46&0.017&0.017&0.018&0.018&0.017&0.017&0.017&0.018&0.018\\
47&0.017&0.018&0.018&0.018&0.018&0.017&0.017&0.018&0.018\\
48&0.017&0.017&0.018&0.018&0.017&0.017&0.017&0.017&0.017\\
49&0.017&0.017&0.017&0.017&0.017&0.017&0.017&0.017&0.018\\
50&0.016&0.016&0.016&0.015&0.015&0.016&0.016&0.016&0.016\\
51&0.017&0.017&0.018&0.018&0.018&0.018&0.018&0.018&0.018\\
52&0.017&0.017&0.018&0.018&0.018&0.018&0.018&0.017&0.018\\
53&0.017&0.017&0.017&0.017&0.017&0.017&0.017&0.017&0.018\\
54&0.017&0.017&0.017&0.017&0.017&0.017&0.017&0.018&0.018\\
55&0.016&0.016&0.016&0.016&0.016&0.016&0.016&0.016&0.016\\
56&0.017&0.017&0.017&0.017&0.017&0.017&0.017&0.017&0.017\\
57&0.017&0.017&0.017&0.017&0.017&0.017&0.017&0.017&0.018\\
 \hline \hline
\end{tabular}}
\label{table:Growth Weights AIC}
\end{table}

\begin{table}[H]
\center
\caption{Weight allocation by BIC for quantiles of interest GDP Growth}
\vspace{0.5cm}
\footnotesize{\begin{tabular}{p{0.1\textwidth}lllllllll}
\hline
\hline
Quantiles Model ID &0.1&0.2&0.3&0.4&0.5&0.6&0.7&0.8&0.9\\
\hline
1&0.021&0.022&0.022&0.022&0.022&0.022&0.022&0.022&0.022\\
2&0.02&0.021&0.021&0.021&0.021&0.021&0.021&0.021&0.02\\
3&0.02&0.02&0.02&0.02&0.021&0.021&0.02&0.02&0.02\\
4&0.018&0.019&0.019&0.019&0.019&0.019&0.019&0.019&0.02\\
5&0.018&0.019&0.019&0.019&0.019&0.019&0.019&0.019&0.019\\
6&0.021&0.021&0.022&0.022&0.022&0.022&0.021&0.021&0.02\\
7&0.021&0.021&0.021&0.021&0.021&0.021&0.021&0.021&0.021\\
8&0.021&0.021&0.021&0.021&0.021&0.021&0.021&0.021&0.02\\
9&0.02&0.021&0.021&0.021&0.021&0.021&0.021&0.02&0.019\\
10&0.019&0.019&0.019&0.019&0.019&0.019&0.019&0.019&0.019\\
11&0.019&0.019&0.019&0.019&0.019&0.019&0.018&0.018&0.018\\
12&0.019&0.019&0.019&0.019&0.019&0.018&0.018&0.018&0.017\\
13&0.017&0.016&0.016&0.016&0.016&0.016&0.016&0.016&0.016\\
14&0.016&0.016&0.015&0.015&0.015&0.015&0.015&0.015&0.016\\
15&0.016&0.016&0.015&0.014&0.013&0.013&0.012&0.011&0.01\\
16&0.02&0.02&0.02&0.02&0.02&0.02&0.021&0.021&0.02\\
17&0.02&0.02&0.02&0.02&0.02&0.02&0.02&0.02&0.02\\
18&0.019&0.019&0.02&0.02&0.02&0.02&0.02&0.02&0.02\\
19&0.019&0.019&0.019&0.02&0.02&0.02&0.02&0.02&0.02\\
20&0.018&0.019&0.018&0.018&0.019&0.019&0.019&0.019&0.018\\
21&0.018&0.018&0.018&0.018&0.018&0.018&0.018&0.018&0.018\\
22&0.018&0.018&0.018&0.018&0.018&0.018&0.018&0.018&0.018\\
23&0.018&0.018&0.018&0.018&0.018&0.018&0.018&0.018&0.018\\
24&0.018&0.018&0.018&0.018&0.018&0.018&0.018&0.018&0.018\\
25&0.016&0.016&0.016&0.016&0.017&0.017&0.017&0.017&0.017\\
26&0.019&0.019&0.019&0.019&0.019&0.019&0.019&0.019&0.019\\
27&0.019&0.019&0.019&0.019&0.019&0.019&0.019&0.018&0.018\\
28&0.018&0.019&0.019&0.019&0.019&0.019&0.018&0.018&0.017\\
29&0.018&0.018&0.018&0.018&0.018&0.018&0.018&0.019&0.019\\
30&0.018&0.018&0.018&0.018&0.018&0.018&0.018&0.018&0.019\\
31&0.018&0.018&0.018&0.018&0.018&0.018&0.018&0.018&0.018\\
32&0.017&0.017&0.016&0.016&0.016&0.016&0.016&0.016&0.017\\
33&0.016&0.017&0.016&0.016&0.016&0.016&0.016&0.016&0.017\\
34&0.016&0.016&0.016&0.016&0.016&0.016&0.015&0.015&0.015\\
35&0.014&0.014&0.013&0.013&0.013&0.013&0.013&0.013&0.013\\
36&0.018&0.018&0.018&0.018&0.018&0.018&0.018&0.018&0.018\\
37&0.017&0.018&0.018&0.018&0.018&0.018&0.018&0.018&0.018\\
38&0.017&0.017&0.018&0.018&0.018&0.018&0.018&0.018&0.018\\
39&0.017&0.017&0.017&0.017&0.017&0.017&0.018&0.018&0.018\\
40&0.017&0.017&0.017&0.017&0.017&0.017&0.017&0.018&0.018\\
41&0.017&0.017&0.017&0.017&0.017&0.017&0.017&0.017&0.018\\
42&0.016&0.016&0.016&0.016&0.016&0.016&0.016&0.016&0.016\\
43&0.016&0.016&0.015&0.015&0.015&0.016&0.016&0.016&0.016\\
44&0.015&0.015&0.015&0.015&0.015&0.015&0.015&0.015&0.016\\
45&0.016&0.015&0.015&0.015&0.015&0.015&0.015&0.015&0.016\\
46&0.016&0.016&0.017&0.016&0.016&0.016&0.016&0.016&0.017\\
47&0.016&0.016&0.016&0.016&0.016&0.016&0.016&0.016&0.017\\
48&0.016&0.016&0.016&0.016&0.016&0.016&0.016&0.016&0.016\\
49&0.016&0.016&0.016&0.016&0.016&0.016&0.016&0.016&0.016\\
50&0.015&0.014&0.014&0.014&0.014&0.014&0.014&0.014&0.015\\
51&0.016&0.016&0.016&0.016&0.016&0.016&0.016&0.016&0.017\\
52&0.016&0.016&0.016&0.016&0.016&0.016&0.016&0.016&0.016\\
53&0.016&0.016&0.016&0.016&0.016&0.016&0.016&0.016&0.016\\
54&0.015&0.015&0.015&0.015&0.015&0.015&0.016&0.016&0.016\\
55&0.015&0.014&0.014&0.014&0.014&0.014&0.014&0.014&0.015\\
56&0.015&0.015&0.015&0.015&0.015&0.015&0.015&0.015&0.016\\
57&0.015&0.015&0.015&0.015&0.015&0.015&0.016&0.016&0.016\\
 \hline \hline
\end{tabular}}
\label{table:Growth Weights BIC}
\end{table}

\begin{table}[H]
\center
\caption{Weight allocation by QRIC for quantiles of interest GDP Growth}
\vspace{0.5cm}
\footnotesize{\begin{tabular}{p{0.1\textwidth}lllllllll}
\hline
\hline
Quantiles Model ID &0.1&0.2&0.3&0.4&0.5&0.6&0.7&0.8&0.9\\
\hline
1&0&0&0&0&0&0&0&0&0\\
2&0&0&0&0&0&0&0&0&0\\
3&0.049&0&0.037&0.102&0.102&0&0.139&0.104&0.177\\
4&0&0&0&0&0&0&0&0.137&0.051\\
5&0&0&0&0&0.009&0&0&0&0\\
6&0.937&0.769&0.624&0.157&0.536&0.671&0.386&0.547&0.234\\
7&0&0&0&0&0&0&0&0&0\\
8&0&0&0&0&0&0&0.336&0.149&0.473\\
9&0&0&0.057&0&0&0&0&0&0\\
10&0&0&0&0&0&0&0&0&0\\
11&0&0&0&0&0&0&0&0&0\\
12&0&0&0&0&0&0&0&0&0\\
13&0&0&0&0&0&0&0&0&0\\
14&0&0&0&0&0&0.013&0.016&0&0\\
15&0&0&0&0&0&0&0&0.022&0.065\\
16&0&0&0&0&0&0&0&0&0\\
17&0&0&0&0&0&0&0&0&0\\
18&0&0&0&0&0&0&0&0&0\\
19&0&0&0&0&0&0&0&0&0\\
20&0&0&0&0&0&0&0&0&0\\
21&0&0&0&0&0&0&0&0&0\\
22&0&0&0&0&0&0&0&0&0\\
23&0.013&0.231&0&0&0.161&0&0&0.041&0\\
24&0&0&0&0&0&0&0&0&0\\
25&0&0&0&0&0&0&0&0&0\\
26&0&0&0&0&0&0&0&0&0\\
27&0&0&0&0&0&0&0&0&0\\
28&0&0&0&0.648&0.192&0&0&0&0\\
29&0&0&0&0&0&0&0&0&0\\
30&0&0&0&0&0&0&0&0&0\\
31&0&0&0&0&0&0&0&0&0\\
32&0&0&0&0&0&0&0&0&0\\
33&0&0&0&0&0&0&0&0&0\\
34&0&0&0&0&0&0&0&0&0\\
35&0&0&0&0&0&0&0&0&0\\
36&0&0&0&0&0&0&0&0&0\\
37&0&0&0&0&0&0&0&0&0\\
38&0&0&0&0&0&0&0&0&0\\
39&0&0&0&0&0&0&0&0&0\\
40&0&0&0&0&0&0&0&0&0\\
41&0&0&0&0&0&0&0&0&0\\
42&0&0&0&0&0&0&0&0&0\\
43&0&0&0&0&0&0&0&0&0\\
44&0&0&0&0&0&0&0&0&0\\
45&0&0&0&0.093&0&0.316&0.123&0&0\\
46&0&0&0&0&0&0&0&0&0\\
47&0&0&0&0&0&0&0&0&0\\
48&0&0&0&0&0&0&0&0&0\\
49&0&0&0&0&0&0&0&0&0\\
50&0&0&0&0&0&0&0&0&0\\
51&0&0&0.012&0&0&0&0&0&0\\
52&0&0&0&0&0&0&0&0&0\\
53&0&0&0&0&0&0&0&0&0\\
54&0&0&0&0&0&0&0&0&0\\
55&0&0&0&0&0&0&0&0&0\\
56&0&0&0&0&0&0&0&0&0\\
57&0&0&0.27&0&0&0&0&0&0\\
 \hline \hline
\end{tabular}}
\label{table:Growth Weights QRIC}
\end{table}

\begin{table}[H]
\center
\caption{Weight allocation by Cross-Validation for quantiles of interest for GDP Growth}
\vspace{0.5cm}
\footnotesize{\begin{tabular}{p{0.1\textwidth}lllllllll}
\hline
\hline
Quantiles Model ID &0.1&0.2&0.3&0.4&0.5&0.6&0.7&0.8&0.9\\
\hline
1&0&0&0&0&0&0&0&0&0\\
2&0&0&0&0&0&0&0&0&0\\
3&0&0&0&0&0&0&0&0.002&0\\
4&0&0&0&0&0&0&0&0&0\\
5&0&0.007&0&0&0&0&0&0.001&0\\
6&0&0&0&0&0&0&0&0&0\\
7&0&0&0&0&0&0&0&0&0\\
8&0&0&0&0&0&0&0&0&0\\
9&0&0&0&0&0&0&0&0&0\\
10&0&0&0&0&0&0&0&0&0\\
11&0&0&0&0&0&0&0&0&0\\
12&0&0&0&0&0&0.006&0&0&0\\
13&0&0&0&0&0&0&0&0&0\\
14&0.003&0&0&0&0&0&0.001&0&0.015\\
15&0&0&0&0&0&0&0&0.008&0.033\\
16&0&0&0&0&0&0&0&0&0\\
17&0&0&0&0&0&0&0&0&0\\
18&0&0&0&0&0&0&0&0&0\\
19&0&0&0&0&0&0&0&0&0\\
20&0.031&0&0&0&0&0&0&0&0\\
21&0&0&0&0&0&0&0&0&0\\
22&0&0&0&0&0&0&0&0&0\\
23&0.001&0.001&0&0&0&0&0&0&0.083\\
24&0&0&0&0.022&0&0.001&0&0.002&0\\
25&0&0&0&0&0&0&0&0&0\\
26&0&0&0&0&0&0&0&0&0\\
27&0&0&0&0&0&0&0&0.247&0\\
28&0&0&0&0.083&0.023&0&0&0&0.154\\
29&0&0&0&0&0&0&0&0&0\\
30&0&0&0&0&0&0&0&0&0\\
31&0&0&0&0&0&0&0.01&0&0.057\\
32&0&0&0&0&0&0&0&0&0\\
33&0&0&0&0&0&0&0&0&0\\
34&0&0&0&0&0&0&0&0&0\\
35&0&0&0&0&0&0&0&0.008&0\\
36&0.443&0&0&0&0&0&0&0.004&0\\
37&0&0&0&0&0&0&0&0&0\\
38&0&0&0&0.005&0&0&0&0&0\\
39&0&0&0&0&0&0&0&0&0\\
40&0&0&0&0.028&0&0.202&0&0&0\\
41&0&0&0&0&0&0&0&0&0\\
42&0&0&0&0&0&0&0&0&0\\
43&0&0&0&0&0&0&0&0&0\\
44&0&0&0&0&0&0&0&0&0\\
45&0&0&0&0&0&0&0.046&0.097&0\\
46&0&0&0&0&0&0&0&0&0\\
47&0&0&0&0&0&0&0&0&0\\
48&0&0&0.285&0.184&0.001&0&0.106&0&0.144\\
49&0&0&0&0&0&0&0&0&0\\
50&0&0&0&0&0&0&0&0&0\\
51&0.522&0&0&0&0&0&0&0&0.264\\
52&0&0&0&0.283&0&0.007&0&0.03&0\\
53&0&0&0&0.007&0&0&0&0&0\\
54&0&0&0&0&0&0&0&0&0\\
55&0&0&0&0&0&0&0.025&0&0\\
56&0&0&0&0.024&0&0&0&0&0\\
57&0&0.992&0.715&0.364&0.977&0.785&0.812&0.602&0.248\\
 \hline \hline
\end{tabular}}
\label{table:Growth Weights Jackknife}
\end{table}

\begin{table}[H]
\center
\caption{Weight allocation by AIC for quantiles of interest for CPI Inlfation}
\vspace{0.5cm}
\footnotesize{\begin{tabular}{p{0.1\textwidth}lllllllll}
\hline
\hline
Quantiles Model ID &0.1&0.2&0.3&0.4&0.5&0.6&0.7&0.8&0.9\\
\hline
1&0.019&0.02&0.02&0.02&0.02&0.02&0.02&0.02&0.02\\
2&0.019&0.019&0.019&0.019&0.019&0.019&0.019&0.019&0.019\\
3&0.018&0.018&0.018&0.018&0.017&0.017&0.017&0.017&0.017\\
4&0.018&0.018&0.018&0.018&0.018&0.018&0.017&0.017&0.016\\
5&0.017&0.017&0.017&0.017&0.017&0.016&0.016&0.016&0.016\\
6&0.019&0.019&0.019&0.019&0.02&0.019&0.019&0.019&0.019\\
7&0.019&0.019&0.019&0.019&0.019&0.02&0.02&0.02&0.02\\
8&0.019&0.019&0.019&0.019&0.019&0.019&0.019&0.019&0.019\\
9&0.019&0.019&0.019&0.019&0.019&0.019&0.019&0.019&0.019\\
10&0.018&0.018&0.018&0.018&0.018&0.018&0.018&0.018&0.019\\
11&0.018&0.018&0.018&0.018&0.018&0.018&0.018&0.018&0.018\\
12&0.018&0.018&0.018&0.018&0.018&0.018&0.018&0.018&0.018\\
13&0.017&0.016&0.016&0.016&0.015&0.015&0.015&0.014&0.014\\
14&0.016&0.016&0.016&0.016&0.015&0.015&0.014&0.014&0.013\\
15&0.016&0.015&0.014&0.014&0.013&0.013&0.012&0.011&0.01\\
16&0.018&0.019&0.019&0.019&0.019&0.019&0.019&0.019&0.019\\
17&0.018&0.019&0.019&0.019&0.019&0.019&0.019&0.019&0.019\\
18&0.018&0.019&0.019&0.019&0.019&0.019&0.019&0.019&0.019\\
19&0.019&0.019&0.019&0.019&0.019&0.019&0.019&0.019&0.019\\
20&0.018&0.018&0.018&0.018&0.018&0.018&0.018&0.018&0.018\\
21&0.018&0.018&0.018&0.018&0.018&0.018&0.018&0.018&0.019\\
22&0.018&0.018&0.018&0.018&0.018&0.018&0.018&0.018&0.018\\
23&0.017&0.017&0.017&0.017&0.017&0.017&0.017&0.017&0.017\\
24&0.017&0.017&0.016&0.016&0.016&0.016&0.016&0.016&0.016\\
25&0.017&0.017&0.017&0.017&0.017&0.016&0.016&0.016&0.015\\
26&0.018&0.018&0.018&0.018&0.018&0.019&0.019&0.019&0.019\\
27&0.018&0.018&0.018&0.018&0.018&0.018&0.018&0.018&0.018\\
28&0.018&0.018&0.018&0.018&0.018&0.018&0.018&0.018&0.018\\
29&0.018&0.018&0.018&0.018&0.018&0.018&0.018&0.018&0.019\\
30&0.018&0.018&0.018&0.018&0.018&0.018&0.019&0.019&0.019\\
31&0.018&0.018&0.018&0.018&0.018&0.018&0.018&0.018&0.018\\
32&0.017&0.017&0.017&0.017&0.017&0.017&0.017&0.017&0.018\\
33&0.017&0.017&0.017&0.017&0.017&0.017&0.017&0.017&0.018\\
34&0.017&0.017&0.017&0.017&0.017&0.017&0.017&0.017&0.017\\
35&0.016&0.015&0.015&0.015&0.014&0.014&0.014&0.013&0.013\\
36&0.017&0.018&0.018&0.018&0.018&0.018&0.018&0.018&0.018\\
37&0.018&0.018&0.018&0.018&0.018&0.018&0.018&0.018&0.018\\
38&0.018&0.018&0.018&0.018&0.018&0.018&0.018&0.018&0.018\\
39&0.018&0.018&0.018&0.018&0.018&0.018&0.018&0.018&0.018\\
40&0.018&0.018&0.018&0.018&0.018&0.018&0.018&0.018&0.018\\
41&0.018&0.018&0.018&0.018&0.018&0.018&0.018&0.018&0.018\\
42&0.017&0.017&0.017&0.017&0.017&0.017&0.017&0.017&0.018\\
43&0.017&0.017&0.017&0.017&0.017&0.017&0.017&0.017&0.017\\
44&0.017&0.017&0.017&0.017&0.017&0.017&0.017&0.017&0.017\\
45&0.016&0.016&0.016&0.016&0.016&0.016&0.016&0.016&0.016\\
46&0.017&0.017&0.017&0.017&0.017&0.017&0.017&0.018&0.018\\
47&0.017&0.017&0.017&0.017&0.017&0.018&0.018&0.018&0.018\\
48&0.017&0.017&0.017&0.017&0.017&0.017&0.017&0.017&0.017\\
49&0.017&0.017&0.017&0.017&0.017&0.017&0.018&0.018&0.018\\
50&0.016&0.016&0.016&0.016&0.016&0.016&0.016&0.016&0.017\\
51&0.017&0.017&0.017&0.017&0.017&0.017&0.017&0.017&0.018\\
52&0.017&0.017&0.017&0.017&0.017&0.017&0.017&0.017&0.018\\
53&0.017&0.017&0.017&0.017&0.017&0.017&0.017&0.017&0.018\\
54&0.017&0.017&0.017&0.017&0.017&0.017&0.017&0.017&0.018\\
55&0.016&0.016&0.016&0.016&0.016&0.016&0.016&0.017&0.017\\
56&0.017&0.017&0.017&0.017&0.017&0.017&0.017&0.017&0.018\\
57&0.017&0.017&0.017&0.017&0.017&0.017&0.017&0.017&0.017\\
 \hline \hline
\end{tabular}}
\label{table:Inflation Weights AIC}
\end{table}

\begin{table}[H]
\center
\caption{Weight allocation by BIC for quantiles of interest for CPI Inlfation}
\vspace{0.5cm}
\footnotesize{\begin{tabular}{p{0.1\textwidth}lllllllll}
\hline
\hline
Quantiles Model ID &0.1&0.2&0.3&0.4&0.5&0.6&0.7&0.8&0.9\\
\hline
1&0.021&0.021&0.022&0.022&0.022&0.022&0.022&0.022&0.022\\
2&0.02&0.021&0.021&0.021&0.021&0.021&0.021&0.021&0.021\\
3&0.019&0.019&0.019&0.019&0.019&0.019&0.019&0.019&0.019\\
4&0.019&0.019&0.019&0.019&0.019&0.019&0.019&0.018&0.018\\
5&0.019&0.019&0.019&0.018&0.018&0.018&0.018&0.017&0.017\\
6&0.02&0.021&0.021&0.021&0.021&0.021&0.021&0.021&0.02\\
7&0.02&0.021&0.021&0.021&0.021&0.021&0.021&0.021&0.021\\
8&0.02&0.02&0.021&0.021&0.021&0.021&0.021&0.021&0.02\\
9&0.02&0.02&0.021&0.021&0.021&0.021&0.021&0.021&0.02\\
10&0.019&0.02&0.02&0.02&0.02&0.02&0.02&0.02&0.02\\
11&0.019&0.019&0.019&0.019&0.019&0.019&0.019&0.019&0.019\\
12&0.019&0.019&0.019&0.019&0.019&0.019&0.019&0.019&0.019\\
13&0.017&0.017&0.017&0.017&0.016&0.016&0.015&0.015&0.014\\
14&0.017&0.017&0.017&0.016&0.016&0.015&0.015&0.014&0.013\\
15&0.016&0.015&0.015&0.014&0.014&0.013&0.012&0.011&0.01\\
16&0.019&0.02&0.02&0.02&0.02&0.02&0.02&0.02&0.02\\
17&0.019&0.02&0.02&0.02&0.02&0.02&0.02&0.02&0.02\\
18&0.019&0.019&0.02&0.02&0.02&0.02&0.02&0.02&0.02\\
19&0.019&0.02&0.02&0.02&0.02&0.02&0.02&0.02&0.02\\
20&0.018&0.019&0.019&0.019&0.019&0.019&0.019&0.019&0.019\\
21&0.018&0.018&0.018&0.019&0.019&0.019&0.019&0.019&0.019\\
22&0.018&0.018&0.018&0.018&0.018&0.018&0.019&0.019&0.019\\
23&0.018&0.018&0.018&0.018&0.018&0.018&0.017&0.017&0.017\\
24&0.017&0.017&0.017&0.017&0.016&0.016&0.016&0.016&0.016\\
25&0.017&0.017&0.017&0.017&0.017&0.017&0.016&0.016&0.015\\
26&0.018&0.018&0.019&0.019&0.019&0.019&0.019&0.019&0.019\\
27&0.018&0.018&0.018&0.019&0.019&0.019&0.018&0.018&0.018\\
28&0.018&0.018&0.018&0.018&0.019&0.019&0.018&0.018&0.018\\
29&0.018&0.018&0.018&0.018&0.018&0.018&0.018&0.019&0.019\\
30&0.018&0.018&0.018&0.018&0.018&0.018&0.019&0.019&0.019\\
31&0.018&0.018&0.018&0.018&0.018&0.018&0.018&0.018&0.018\\
32&0.017&0.017&0.017&0.017&0.017&0.017&0.017&0.017&0.017\\
33&0.017&0.017&0.017&0.017&0.017&0.017&0.017&0.017&0.018\\
34&0.017&0.016&0.016&0.017&0.017&0.017&0.016&0.016&0.016\\
35&0.015&0.015&0.014&0.014&0.014&0.013&0.013&0.012&0.012\\
36&0.017&0.017&0.017&0.017&0.018&0.018&0.018&0.018&0.018\\
37&0.017&0.017&0.017&0.017&0.017&0.017&0.018&0.018&0.018\\
38&0.017&0.017&0.017&0.017&0.017&0.018&0.018&0.018&0.018\\
39&0.017&0.017&0.017&0.017&0.017&0.017&0.017&0.018&0.018\\
40&0.017&0.017&0.017&0.017&0.017&0.017&0.017&0.017&0.018\\
41&0.017&0.017&0.017&0.017&0.017&0.017&0.017&0.017&0.018\\
42&0.016&0.016&0.016&0.016&0.016&0.016&0.016&0.017&0.017\\
43&0.016&0.016&0.016&0.016&0.016&0.016&0.016&0.016&0.016\\
44&0.016&0.016&0.016&0.016&0.016&0.016&0.016&0.016&0.017\\
45&0.016&0.015&0.015&0.015&0.015&0.015&0.015&0.015&0.015\\
46&0.016&0.016&0.016&0.016&0.016&0.016&0.016&0.017&0.017\\
47&0.016&0.016&0.016&0.016&0.016&0.016&0.017&0.017&0.017\\
48&0.016&0.016&0.016&0.016&0.016&0.016&0.016&0.016&0.016\\
49&0.016&0.016&0.016&0.016&0.016&0.016&0.016&0.016&0.017\\
50&0.015&0.015&0.015&0.015&0.015&0.015&0.015&0.015&0.016\\
51&0.016&0.016&0.016&0.016&0.016&0.016&0.016&0.016&0.016\\
52&0.016&0.016&0.016&0.016&0.016&0.016&0.016&0.016&0.016\\
53&0.016&0.016&0.016&0.016&0.016&0.016&0.016&0.016&0.016\\
54&0.016&0.016&0.015&0.015&0.015&0.016&0.016&0.016&0.016\\
55&0.015&0.015&0.015&0.015&0.014&0.015&0.015&0.015&0.016\\
56&0.015&0.015&0.015&0.015&0.015&0.015&0.015&0.016&0.016\\
57&0.015&0.015&0.015&0.015&0.015&0.015&0.015&0.016&0.016\\
 \hline \hline
\end{tabular}}
\label{table:Inflation Weights BIC}
\end{table}

\begin{table}[H]
\center
\caption{Weight allocation by QRIC for quantiles of interest for CPI Inlfation}
\vspace{0.5cm}
\footnotesize{\begin{tabular}{p{0.1\textwidth}lllllllll}
\hline
\hline
Quantiles Model ID &0.1&0.2&0.3&0.4&0.5&0.6&0.7&0.8&0.9\\
\hline
1&0&0.178&0.081&0&0&0.439&0.002&0&0\\
2&0&0&0&0&0&0&0&0&0\\
3&0&0.019&0&0.117&0.155&0.047&0.058&0.013&0.096\\
4&0&0&0&0.036&0.059&0.036&0.055&0.133&0.094\\
5&0&0&0.013&0.095&0&0&0.004&0&0\\
6&0&0&0.207&0.021&0.384&0.029&0.014&0&0\\
7&0&0&0&0&0&0&0.033&0&0\\
8&0&0&0.048&0&0&0&0.036&0&0\\
9&0.703&0&0.13&0.73&0.402&0.385&0.771&0.844&0.435\\
10&0&0&0&0&0&0&0&0&0\\
11&0&0&0&0&0&0&0&0&0\\
12&0&0.156&0&0&0&0&0&0&0.263\\
13&0&0&0&0&0&0&0.025&0&0\\
14&0&0&0&0&0&0&0.001&0&0.05\\
15&0&0&0&0&0&0&0&0.002&0.062\\
16&0&0&0&0&0&0&0&0&0\\
17&0&0&0&0&0&0&0&0&0\\
18&0&0&0&0&0&0&0&0&0\\
19&0&0.455&0.476&0&0&0&0&0&0\\
20&0&0&0&0&0&0&0&0&0\\
21&0&0&0&0&0&0&0&0&0\\
22&0&0&0&0&0&0&0&0&0\\
23&0.297&0.009&0.004&0&0&0.001&0&0&0\\
24&0&0&0&0&0&0&0&0&0\\
25&0&0.183&0.04&0&0&0&0&0&0\\
26&0&0&0&0&0&0&0&0&0\\
27&0&0&0&0&0&0&0&0&0\\
28&0&0&0&0&0&0.063&0&0&0\\
29&0&0&0&0&0&0&0&0&0\\
30&0&0&0&0&0&0&0&0&0\\
31&0&0&0&0&0&0&0&0&0\\
32&0&0&0&0&0&0&0&0&0\\
33&0&0&0&0&0&0&0&0&0\\
34&0&0&0&0&0&0&0&0&0\\
35&0&0&0&0&0&0&0&0.008&0\\
36&0&0&0&0&0&0&0&0&0\\
37&0&0&0&0&0&0&0&0&0\\
38&0&0&0&0&0&0&0&0&0\\
39&0&0&0&0&0&0&0&0&0\\
40&0&0&0&0&0&0&0&0&0\\
41&0&0&0&0&0&0&0&0&0\\
42&0&0&0&0&0&0&0&0&0\\
43&0&0&0&0&0&0&0&0&0\\
44&0&0&0&0&0&0&0&0&0\\
45&0&0&0&0&0&0&0&0&0\\
46&0&0&0&0&0&0&0&0&0\\
47&0&0&0&0&0&0&0&0&0\\
48&0&0&0&0&0&0&0&0&0\\
49&0&0&0&0&0&0&0&0&0\\
50&0&0&0&0&0&0&0&0&0\\
51&0&0&0&0&0&0&0&0&0\\
52&0&0&0&0&0&0&0&0&0\\
53&0&0&0&0&0&0&0&0&0\\
54&0&0&0&0&0&0&0&0&0\\
55&0&0&0&0&0&0&0&0&0\\
56&0&0&0&0&0&0&0&0&0\\
57&0&0&0&0&0&0&0&0&0\\
 \hline \hline
\end{tabular}}
\label{table:Inflation Weights QRIC}
\end{table}

\begin{table}[H]
\center
\caption{Weight allocation by Cross-Validation for quantiles of interest for CPI Inlfation}
\vspace{0.5cm}
\footnotesize{\begin{tabular}{p{0.1\textwidth}lllllllll}
\hline
\hline
Quantiles Model ID &0.1&0.2&0.3&0.4&0.5&0.6&0.7&0.8&0.9\\
\hline
1&0&0&0&0&0&0.006&0&0&0\\
2&0&0&0&0&0&0&0&0&0\\
3&0&0&0&0.022&0.003&0.049&0&0&0.028\\
4&0&0&0&0&0&0&0&0&0\\
5&0&0&0&0&0&0&0.002&0&0\\
6&0&0&0&0&0&0&0&0&0\\
7&0&0&0&0&0&0&0&0&0\\
8&0&0&0&0&0&0&0&0&0\\
9&0&0&0&0.172&0&0&0&0&0.073\\
10&0&0&0&0&0&0&0&0&0\\
11&0&0&0&0&0&0&0&0&0.01\\
12&0&0.005&0&0&0&0&0&0&0.113\\
13&0&0&0&0&0&0&0.002&0&0\\
14&0&0&0&0&0&0&0&0&0.002\\
15&0&0&0&0&0&0&0&0&0.065\\
16&0&0&0&0&0&0.032&0&0&0\\
17&0&0&0&0.055&0&0.01&0&0&0\\
18&0&0&0&0&0&0&0&0&0\\
19&0&0.032&0&0&0&0&0&0&0\\
20&0&0&0&0&0&0.007&0&0&0\\
21&0&0&0&0&0&0&0&0&0\\
22&0&0&0&0&0&0&0&0&0\\
23&0&0.001&0.012&0.029&0.01&0&0&0&0.042\\
24&0&0&0.011&0.062&0&0&0&0&0\\
25&0&0.056&0&0&0&0&0.005&0.002&0\\
26&0&0&0&0&0&0&0&0&0\\
27&0&0&0.048&0&0&0&0&0&0\\
28&0&0.013&0&0.041&0.128&0.158&0.104&0.553&0.082\\
29&0&0&0&0&0&0&0&0&0\\
30&0&0.024&0&0&0&0&0&0&0.038\\
31&0.319&0&0.002&0&0&0&0&0&0.029\\
32&0&0&0&0&0&0&0&0&0\\
33&0&0&0&0&0&0&0&0&0.004\\
34&0&0&0&0&0&0&0&0&0.098\\
35&0&0&0&0&0&0&0&0.05&0.042\\
36&0&0&0&0&0&0.011&0&0&0\\
37&0&0&0&0&0&0&0.001&0&0\\
38&0&0&0&0.102&0&0.122&0&0&0\\
39&0&0&0&0.051&0&0&0&0&0\\
40&0&0.042&0&0.007&0&0&0&0&0\\
41&0&0.108&0&0.005&0&0&0&0&0.031\\
42&0&0&0&0&0&0&0&0&0\\
43&0&0&0&0&0&0&0&0&0\\
44&0&0&0&0&0&0&0&0.004&0.001\\
45&0.127&0.074&0&0&0.107&0&0&0.01&0.048\\
46&0&0&0&0&0&0&0&0&0\\
47&0&0.041&0&0&0&0.057&0&0.046&0.04\\
48&0&0&0&0.022&0.276&0.062&0&0&0.022\\
49&0&0.048&0&0&0&0&0&0&0.037\\
50&0&0&0&0&0&0&0&0&0.016\\
51&0&0&0.086&0&0&0.034&0&0&0\\
52&0&0&0&0.059&0&0.122&0&0&0.004\\
53&0&0.121&0&0.262&0.001&0.122&0.278&0.096&0.043\\
54&0&0.091&0&0.003&0&0.009&0&0&0.033\\
55&0&0.211&0&0&0&0&0&0&0.016\\
56&0&0.056&0&0&0&0.064&0.23&0&0.042\\
57&0.554&0.074&0.842&0.107&0.476&0.136&0.377&0.239&0.043\\
 \hline \hline
\end{tabular}}
\label{table:Inflation Weights Jackknife}
\end{table}

\section{Forecast Performance Measures}
\begin{table}[H]
\center
\caption{Empirical Coverage Rates for one-quarter-ahead forecasts of GDP growth}
\vspace{0.5cm}
\begin{threeparttable}
\begin{tabular}{cp{0.075\textwidth}p{0.075\textwidth}p{0.075\textwidth}p{0.075\textwidth}p{0.075\textwidth}p{0.075\textwidth}p{0.1\textwidth}}
\hline
\hline
\textbf{\begin{tabular}[c]{@{}c@{}}Quantile\\ ($\tau$)\end{tabular}} & \textbf{QAR(1)} & \textbf{Full Model} & \textbf{Equal} & \textbf{AIC} & \textbf{BIC} & \textbf{QRIC} & \textbf{Jackknife} \\
\hline
\textbf{0.10}                                                      & 0.0566         & $0.2075^{**}$              & 0.0566         & 0.0566       & 0.0566       & 0.0755        & $0.1698^*$             \\
\textbf{0.20}                                                      & $0.1698^*$           & 0.1321              & 0.1321         & 0.1321       & 0.1321       & 0.1509        & $0.1887^{**}$             \\
\textbf{0.30}                                                      & $0.2453^{**}$           & 0.0377              & $0.2642^{**}$          & $0.2642^{**}$       & $0.2642^{**}$      & 0.1321        & 0.0566             \\
\textbf{0.40}                                                      & 0.0566          & $0.0189^{**}$               & 0.0377         & 0.0566       & 0.0566       & 0.0566        & 0.0566             \\
\textbf{0.50}                                                      & $0.0000^{**}$          & $0.0189^{**}$                & 0.0755         & 0.0377       & 0.0566       & 0.0943        & $0.0189^{**}$                 \\
\textbf{0.60}                                                      & 0.1132          & 0.0566              & 0.0943         & 0.1132       & 0.0943       & 0.0377        & 0.0566             \\
\textbf{0.70}                                                      & $0.0189^{**}$                & $0.0189^{**}$                   & 0.0566         & 0.0377       & 0.0377       & $0.0189^{**}$  & $0.0000^{**}$                 \\
\textbf{0.80}                                                      & $0.0000^{**}$          & 0.0566              & $0.0189^{**}$    & 0.0377       & 0.0377       & 0.0566        & $0.0189^{**}$                   \\
\textbf{0.90}                                                      & 0.0755          & $0.0189^{**}$                    & 0.0377         & 0.0377       & 0.0377       & $0.0189^{**}$              & 0.0566       \\
 \hline \hline
\end{tabular}
\begin{tablenotes}
            \footnotesize{\item *Significantly different from nominal coverage at 10$\%$ significance level.
            \item **Significantly different from nominal coverage at $5\%$ significance level.}
\end{tablenotes}
\end{threeparttable}
\label{table:GDP Coverage}
\end{table} 

\vspace{0.5cm}

\begin{table}[H]
\center
\caption{Empirical Coverage Rates for one-quarter-ahead forecasts of CPI Inflation}
\vspace{0.5cm}
\begin{threeparttable}
\begin{tabular}{cp{0.075\textwidth}p{0.075\textwidth}p{0.075\textwidth}p{0.075\textwidth}p{0.075\textwidth}p{0.075\textwidth}p{0.1\textwidth}}
\hline
\hline
\textbf{\begin{tabular}[c]{@{}c@{}}Quantile\\ ($\tau$)\end{tabular}} & \textbf{QAR(1)} & \textbf{Full Model} & \textbf{Equal} & \textbf{AIC} & \textbf{BIC} & \textbf{QRIC} & \textbf{Jackknife} \\
\hline
\textbf{0.10}                                                      & 0.1509          & 0.1132              & 0.0755         & 0.0755       & 0.0755       & 0.1509        & 0.0755             \\
\textbf{0.20}                                                      & 0.0566          &  $0.0189^{**}$              & 0.0377         & 0.0377       & 0.0377       & 0.0377        & 0.0377             \\
\textbf{0.30}                                                      & 0.0377          & 0.0566              &  $0.0189^{**}$        &  $0.0189^{**}$        &  $0.0189^{**}$        &  $0.0189^{**}$        & 0.0566             \\
\textbf{0.40}                                                      & 0.1132          & 0.0943              & 0.0566         & 0.0566       & 0.0566       & 0.0943        & 0.0566             \\
\textbf{0.50}                                                      & 0.0377          & 0.0566              & 0.1132         & 0.1132       & 0.1132       & 0.0377        & 0.0943             \\
\textbf{0.60}                                                      & 0.0566          & 0.0377              & 0.0566         & 0.0566       & 0.0377       & 0.0566        & 0.0377             \\
\textbf{0.70}                                                      & 0.1132          & 0.0755              & 0.0377         & 0.0377       & 0.0566       & 0.0566        & 0.0377             \\
\textbf{0.80}                                                      & 0.0755          & 0.0943              & 0.0755         & 0.1132       & 0.1132       & 0.0755        & 0.0566             \\
\textbf{0.90}                                                      & 0.1321          & 0.0755              & $0.1887^{**}$        & $0.1698^{*}$      & $0.1698^{*}$       & $0.1887^{**}$        & 0.1509    \\
 \hline \hline
\end{tabular}
\begin{tablenotes}
            \footnotesize{\item * Significantly different from nominal coverage at 10$\%$ significance level.
            \item **Significantly different from nominal coverage at $5\%$ significance level.}
\end{tablenotes}
\end{threeparttable}
\label{table:CPI Coverage}
\end{table} 

\vspace{0.5cm}

\begin{table}[H]
\center
\caption{Final Prediction Error for one-quarter-ahead forecasts of GDP growth}
\vspace{0.5cm}
\begin{tabular}{cccccccc}
\hline
\hline
\textbf{\begin{tabular}[c]{@{}c@{}}Quantile\\ ($\tau$)\end{tabular}} & \textbf{QAR(1)} & \textbf{Full Model} & \textbf{Equal} & \textbf{AIC} & \textbf{BIC} & \textbf{QRIC} & \textbf{Jackknife} \\
\hline
\textbf{0.10}                                                        & 0.0025          & 0.0033              & 0.0033         & 0.0033       & 0.0033       & 0.0021        & 0.0028             \\
\textbf{0.20}                                                        & 0.0029          & 0.0033              & 0.0034         & 0.0033       & 0.0033       & 0.0025        & 0.0031             \\
\textbf{0.30}                                                        & 0.0033          & 0.0030              & 0.0034         & 0.0034       & 0.0034       & 0.0022        & 0.0025             \\
\textbf{0.40}                                                        & 0.0035          & 0.0034              & 0.0037         & 0.0037       & 0.0037       & 0.0026        & 0.0027             \\
\textbf{0.50}                                                        & 0.0035          & 0.0031              & 0.0038         & 0.0037       & 0.0037       & 0.0028        & 0.0031             \\
\textbf{0.60}                                                        & 0.0032          & 0.0032              & 0.0035         & 0.0035       & 0.0035       & 0.0029        & 0.0031             \\
\textbf{0.70}                                                        & 0.0028          & 0.0027              & 0.0031         & 0.0031       & 0.0031       & 0.0026        & 0.0026             \\
\textbf{0.80}                                                        & 0.0022          & 0.0020              & 0.0025         & 0.0025       & 0.0025       & 0.0019        & 0.0020             \\
\textbf{0.90}                                                        & 0.0016          & 0.0011              & 0.0017         & 0.0017       & 0.0017       & 0.0015        & 0.0013       \\
 \hline \hline
\end{tabular}
\label{table:GDP FPE}
\end{table} 

\vspace{0.5cm}

\begin{table}[H]
\center
\caption{Final Prediction Error for one-quarter-ahead forecasts of CPI Inflation}
\vspace{0.5cm}
\begin{tabular}{cccccccc}
\hline
\hline
\textbf{\begin{tabular}[c]{@{}c@{}}Quantile\\ ($\tau$)\end{tabular}} & \textbf{QAR(1)} & \textbf{Full Model} & \textbf{Equal} & \textbf{AIC} & \textbf{BIC} & \textbf{QRIC} & \textbf{Jackknife} \\
\hline
\textbf{0.10}                                                        & 0.0010          & 0.0012              & 0.0013         & 0.0013       & 0.0013       & 0.0014        & 0.0013             \\
\textbf{0.20}                                                        & 0.0015          & 0.0020              & 0.0020         & 0.0020       & 0.0020       & 0.0020        & 0.0021             \\
\textbf{0.30}                                                        & 0.0019          & 0.0022              & 0.0025         & 0.0024       & 0.0025       & 0.0019        & 0.0022             \\
\textbf{0.40}                                                        & 0.0023          & 0.0022              & 0.0028         & 0.0027       & 0.0027       & 0.0024        & 0.0024             \\
\textbf{0.50}                                                        & 0.0022          & 0.0023              & 0.0029         & 0.0028       & 0.0028       & 0.0024        & 0.0024             \\
\textbf{0.60}                                                        & 0.0021          & 0.0021              & 0.0028         & 0.0028       & 0.0028       & 0.0022        & 0.0022             \\
\textbf{0.70}                                                        & 0.0020          & 0.0021              & 0.0028         & 0.0027       & 0.0027       & 0.0022        & 0.0021             \\
\textbf{0.80}                                                        & 0.0016          & 0.0020              & 0.0025         & 0.0023       & 0.0023       & 0.0020        & 0.0020             \\
\textbf{0.90}                                                        & 0.0012          & 0.0015              & 0.0017         & 0.0016       & 0.0016       & 0.0019        & 0.0019   \\
 \hline \hline
\end{tabular}
\label{table:CPI FPE}
\end{table}

\newpage
\subsection{Dataset}
\begin{longtable}{|l|l|l|}
\hline
\textbf{NR} & \textbf{FAME CODE} & \textbf{Series Name} \\
\hline
1&D7BT&Consumer Price Index: all items\\
2&ABMI&Gross Domestic Product: chained volume measures\\
&&\\
3&ABJR&Household final consumption expenditure \\
4&CKYY&IOP: Industry D: Manufacturing\\
5&CKYZ&IOP: Industry E: Electricity, gas and water supply\\
6&CKZF&IOP: Industry DF: Manufacturing of food, drink and tobacco\\
7&CKZG&IOP: Industry DG: Manufacturing of chemicals and man-made fibres \\
8&GDBQ&ESA95 Output Index: F:Construction\\
9&GDQH&SA95 Output Industry: I: Transport storage and communication\\
10&GDQS&SA95 Output Industry: G-Q: Total\\
11&IKBK&Balance of Payments: Trade in Goods and Services: Total exports\\
12&IKBL&Balance of Payments: Imports: Total Trade in Goods and Services\\
13&NMRY&General Government: Final consumption expenditure\\
14&NPQT&Total Gross Fixed Capital Formation\\ \hline
\multicolumn{3}{c}{ \textbf{Household final consumption expenditure: durable goods}}\\ \hline
15&ATQX&Furniture and households\\
16&ATRD&Carpets and other floor coverings\\
17&ATRR&Telephone and telefax equipment\\
18&ATRV&Audio visual equipment\\
19&ATRZ&Photo and cinema equipment and optical instruments\\
20&ATSD&Information processing equipment\\
21&LLKX&All funrishing and household\\
22&LLKY&All health\\
23&LLKZ&All transport\\
24&LLLA&All communication\\
25&LLLB&All recreation and culture\\
26&LLLC&All miscellaneous\\
27&TMMI&All purchases of vehicles\\
28&TMML&Motor cars\\
29&TMMZ&Motor cycles\\
30&TMNB&Major durables for outdoor recreation\\
31&TMNO&Bicycles\\
32&UTID&Total \\
33&UWIC&Therapeutic appliances and equipment\\
34&XYJP&Major house appliances\\
35&XYJR&Major tools and equipment\\
36&XYJT&Musical instruments and major durables for indoor recreation\\
37&ZAYM&Jewelery, clocks and watches\\ \hline
\multicolumn{3}{c}{ \textbf{Household final consumption expenditure: semi-durable goods}}\\ \hline
38&ATQV&Shoes and other footwear\\
39&ATRF&Household and textiles\\
40&ATRJ&Glassware, tableware and household utensils\\
41&ATSH&Recording media\\
42&ATSL&Games, toys and hobbies\\
43&ATSX&Other personal effects\\
44&AWUW&Motor vehicle spares\\
45&CDZQ&Books\\
46&LLLZ&All clothing and footwear\\
47&LLMC&All recreation and culture\\
48&LLMD&All miscellaneous\\
49&UTIT&Total  \\
50&XYJN&Clothing materials\\
51&XYJO&Other articles of clothing and clothing accessories\\
52&XYJQ&Small eectric household appliances\\
53&XYJS&Small tools and miscellaneous accessories\\
54&XYJU&Equipment for sport, camping etc\\
55&XYJX&Electrical appliances for personal care\\
56&ZAVK&Garments\\ \hline
\multicolumn{3}{c}{ \textbf{Household final consumption expenditure: non-durable goods}}\\ \hline
57&ATSP&Other products for personal care\\
58&ATUA&Materials for the maintenance and repair of the dwelling\\
59&AWUX&Gardens, plants and flowers\\
60&CCTK&Meat\\
61&CCTL&Fish\\
62&CCTM&Milk, cheese and eggs\\
63&CCTN&Oils and fats\\
64&CCTO&Fruit\\
65&CCTT&Coffee,tea and cocoa\\
66&CCTU&Mineral, water and soft drinks\\
67&CCTY&Vehicle fuels and lubricants\\
68&CCUA&Electricity\\
69&CDZY&Newspapers and periodicals\\
70&LLLL&All housing, water, electricity, gas and other fuels\\
71&LLLM&All furnishing and household goods\\
72&LLLN&All health\\
73&LLLO&All transport\\
74&LLLP&All recreation and culture\\
75&LLLQ&All miscellaneous\\
76&LTZA&Gas\\
77&LTZC&Liquid fuels\\
78&TTAB&Solid fuels\\
79&UTHW&Wine, cider and sherry\\
80&UTIL&Total\\
81&UTXP&Pharmaceutical Products\\
82&UTZN&Water supply\\
83&UUIS&Spirits\\
84&UUVG&Beer\\
85&UWBK&All food \\
86&UWBL&Bread and cereals\\
87&UWFD&Vegetables\\
88&UWFX&Sugar and sweet products\\
89&UWGH&Food products n.e.c.\\
90&UWGI&All non-alcoholic beverages\\
91&UWHO&Non-durable household goods\\
92&UWIB&Other medical products\\
93&UWKQ&Pets and related products\\
94&XYJV&Miscellaneous printed matter\\
95&XYJW&Stationary and drawing materials\\
96&ZAKY&All alcoholic beverages and otbacco\\
97&ZWUN&All food and non-alcoholic beverages\\
98&ZWUP&Tobacco\\
99&ZWUR&All elctricity, gas and other fuels\\ \hline
\multicolumn{3}{c}{ \textbf{Household final consumption expenditure: services}}\\ \hline
100&AWUY&Repair and hire of footwear\\
101&AWUZ&Services for the maintenance and reair of the dwelling\\
102&AWVA&Vehicle maintenance and repair\\
103&AWVB&Railways\\
104&AWVC&Air\\
105&AWVD&Sea and inland waterway\\
106&AWVE&Other\\
107&CCUO&Imputed rentals of owner-occupiers\\
108&CCVA&Games of chance\\
109&CCVM&Postal services\\
110&CCVZ&Hairdressing salons and personal grooming establishments\\
111&GBFG&Actual rentals paid by tenants\\
112&GBFK&All imputed rentals for housing\\
113&GBFN&Other imputed rentals\\
114&LLLR&All clothing and footwear\\
115&LLLS&All housing, water, electricity, gas and other fuels\\
116&LLLT&All funrishing and household\\
117&LLLU&All health\\
118&LLLV&Total transport\\
119&LLLW&All communication\\
120&LLLX&All recreation and culture\\
121&LLLY&All miscellaneous\\
122&UTIP&Total\\
123&UTMH&Paramedical services\\
124&UTYF&Hospital services\\
125&UTYH&Life insurance \\
126&UTZX&Sewerage collection\\
127&UWHI&Clothing, repair and hire of clothing\\
128&UWHK&Refuse collection\\
129&UWHM&Repair of furniture, furnishings and floor coverings\\
130&UWHN&Repair of household appliances\\
131&UWIA&Domestic and household services\\
132&UWKO&Repair of audio-visual, ohoto and information processing equipment\\
133&UWKP&Maintenance of other major durables for recreation and culture\\
134&UWLD&Veterinary and other services\\
135&ZAVQ&All actual rentals for housing \\
136&ZAWG&All out-patient services\\
137&ZAWI&Medical services\\
138&ZAWK&Dental services\\
139&ZAWQ&Other vehicle services\\
140&ZAWS&All transport services\\
141&ZAWU&Road\\
142&ZAWY&Telephone and telefax services\\
143&ZAXI&All recreational and cultural services\\
144&ZAXK&Recreational and sporting activities\\
145&ZAXM&Cultural services\\
146&ZAXS&All restaurants and hotels\\
147&ZAXU&All catering services\\
148&ZAXW&Restaurants, cafes etc\\
149&ZAYC&Canteens\\
150&ZAYE&Accommodation services\\
151&ZAYO&Social protection\\
152&ZAYQ&All insurance\\
153&ZAYS&Insurance connected with the dwelling\\
154&ZAYU&Insurance connected with health\\
155&ZAYW&Insurance connected with transport\\
156&ZAZA&All financial services n.e.c.\\
157&ZAZC&All financial services other than FISIM\\
158&ZAZE&Other services n.e.c.\\
159&ZWUT&Education\\ \hline
\multicolumn{3}{c}{ \textbf{Deflators}}\\ \hline
160&ABJS&Consumption\\
161&FRAH&RPI: Total Food\\
162&ROYJ&Wages\\
163&YBGB&GDP Deflator\\ \hline
\multicolumn{3}{c}{ \textbf{Money Series}}\\ \hline
164&M4ISA&M4 Deposits PNFCs\\
165&M4OSA&M4 Deposits OFCs\\
166&M4PSA&M4 Deposits Households\\
167&MALISA&M4 Lending Total\\
168&MALOSA&M4 Lending PNFCs\\
169&MALPSA&M4 Lending Households\\ \hline
\multicolumn{3}{c}{ \textbf{Asset Prices}}\\ \hline
170&&Real nationwide house prices\\
171&GDF Data&FTSE All Share Index\\
172&IMF Data&Nominal Effective Exchange Rate (NEER)\\
173&GDF Data&Pounds to Euro\\
174&GDF Data&Pounds to US dollar\\
175&GDF Data&Pounds to Canadian dollar\\
176&GDF Data&Pounds to Australian dollar\\
\hline
\end{longtable}

\end{document}